\documentclass[aps,preprint]{revtex4}
\usepackage[dvips]{graphicx}
\usepackage{latexsym}
\usepackage{natbib}

\begin{document}
\title{Structure of shells in complex networks}
   
\author{Jia Shao$^{1}$, Sergey V. Buldyrev$^{2,1}$,
    Lidia A. Braunstein$^{3,1}$,\\ 
Shlomo Havlin$^{4}$, and H. Eugene Stanley$^{1}$}

\affiliation{
$^1$Center for Polymer Studies and Department of Physics,
  Boston University, Boston, Massachusetts 02215, USA\\
$^2$Department of Physics, Yeshiva University, 500 West 185th Street, New York, New York 10033, USA\\
$^3$Instituto de Investigaciones F\'isicas de Mar del Plata (IFIMAR)-Departamento
de F\'isica, Facultad de Ciencias Exactas y Naturales, Universidad Nacional de Mar del
Plata-CONICET, Funes 3350, (7600) Mar del Plata, Argentina \\
$^4$Minerva Center and Department of Physics, Bar-Ilan University, 52900
Ramat-Gan, Israel
}

\date{ Revised: \today }

\begin{abstract}
In a network, we define shell $\ell$ as the set of nodes at 
distance $\ell$ with respect to a given node and 
define $r_\ell$ as the fraction of nodes outside shell $\ell$.
In a transport process, information or disease usually diffuses from a random
node and reach nodes shell after shell. 
Thus, understanding the shell structure is crucial for the study of 
the transport property of networks.
We study the statistical properties of the shells from a randomly chosen node.
For a randomly connected network with given degree distribution, we derive analytically the 
degree distribution and average degree of the nodes residing outside shell $\ell$ as a function of $r_\ell$.
Further, we find that $r_\ell$
follows an iterative functional form $r_\ell=\phi(r_{\ell-1})$, where 
$\phi$ is expressed in terms of the generating function of the 
original degree distribution of the network.
Our results can explain the power-law distribution 
of the number of nodes $B_\ell$ found in shells with $\ell$ larger than the 
network diameter $d$, which is the average distance between all pairs of nodes.
For real world networks the theoretical prediction of $r_\ell$
deviates from the empirical $r_\ell$. 
We introduce a network correlation function 
$c(r_\ell)\equiv r_{\ell+1}/\phi(r_\ell)$
to characterize the correlations in the network, where $r_{\ell+1}$ is the
empirical value and $\phi(r_\ell)$ is the theoretical prediction.
$c(r_\ell)=1$ indicates perfect agreement between empirical results and
theory.
We apply $c(r_\ell)$ to several model and real world networks. We find
that the networks fall into two distinct classes: (i) a class of 
{\it poorly-connected} networks with $c(r_\ell)>1$, which have larger average
distances compared with randomly connected networks with the same degree 
distributions;
and (ii) a class of {\it well-connected} networks with 
$c(r_\ell)<1$. Examples of poorly-connected networks include 
the Watts-Strogatz model and networks 
characterizing human collaborations, which
include two citation networks and the actor collaboration network. 
Examples of well-connected networks include the Barab\'{a}si-Albert 
model and the Autonomous System (AS) Internet network.

\end{abstract}

\maketitle
\section{Introduction and recent work}
Many complex systems can be described by networks in which the nodes 
are the elements of the system and the links characterize the 
interactions between the elements. One of 
the most common ways to characterize a network is to determine
its degree distribution.
A classical example of a network is
the Erd\H{o}s-R\'enyi (ER) \cite{er1,bollo} model, 
in which the links are randomly
assigned to randomly selected pairs of nodes. The degree distribution 
of the ER model is characterized by a Poisson 
distribution 
\begin{equation}\label{pER}
P(k)=\exp(-\langle k \rangle)\langle k \rangle^k /k ! ,
\end{equation} 
where $\langle k \rangle $ is the average degree of the network.
Another simple model is a random regular (RR) graph in which each node 
has exactly
$\langle k\rangle=\psi$ links, thus $P(k)=\delta(k-\psi)$. 
The Watts-Strogatz 
model (WS) \cite{ws} is also well-studied, 
where a random fraction $\beta$ of links from a 
regular lattice with $\langle k \rangle =\psi$ are rewired and 
connect any pair of nodes. Changing $\beta$ from 0 to 1, the WS network
interpolates between a regular lattice and an ER graph.
In the last decade, it has been realized that many social, computer, and
biological networks can be approximated by scale-free (SF) models
with a broad degree distribution characterized by a power law 
\begin{equation}\label{pSF}
P(k) \sim k^ {- \lambda},
\end{equation}
with a lower and upper cutoff, $k_{\rm min}$ and $k_{\rm max}$  
\cite{PastorXX,barasci,bararev,mendes,cohena,vespig}. 
A paradigmatic model that explains the
abundance of SF networks in nature is the preferential attachment model of
Barab\'{a}si and Albert (BA)\cite{barasci}.

The degree distribution is not sufficient to characterize the topology 
of a network. Given
a degree distribution, a network can have very different properties such as
clustering and degree-degree correlation. 
For example, the network of movie actors \cite{barasci} in which two actors 
are linked if they play in the same movie, although characterized
by a power-law degree distribution, has higher clustering coefficient 
compared to the SF 
network generated by Molloy-Reed algorithm \cite{MR} with the same 
degree distribution.

Besides the degree distribution and clustering coefficient, 
a network is also characterized by the 
average distance between all pairs of nodes,
which we refer to as the network diameter $d$. 
Random networks with a given degree distribution can be ``small worlds'' \cite{bollo} 
\begin{equation}\label{Eq.dlnn}
d\sim \ln N
\end{equation}
or ``ultra-small worlds'' \cite{cohena}
\begin{equation}\label{Eq.dlnlnn}
d\sim \ln \ln N.
\end{equation}
The diameter $d$ depends sensitively on the network topology.

Another important characteristic of a network is
the structure of its shells, where shell $\ell$
is defined as the set of nodes that are at 
distance $\ell$ from a randomly chosen root node \cite{kalisky}.
The shell structure of a network
is important for understanding the transport properties of 
the network such as the epidemic spread \cite{Newman}, 
where the virus spread from a randomly chosen
root and reach nodes shell after shell. 
The structure of the shells is related to both
the degree distribution and the network diameter. The shell structure
of SF networks has been recently 
studied Ref. \cite{kalisky},
which have introduces a new term ``network tomography'' referring 
to various properties 
of shells such as the number of nodes and open links in shell $\ell$, 
the degree distribution, and the average degree of the nodes in the exterior 
of shell $\ell$. 

Many real and model networks have fractal
properties while others are not \cite{song}. 
Recently Ref. \cite{Shao} reported a power law 
distribution of number of nodes $B_\ell$ in shell $\ell >d$ from a randomly 
chosen root. 
They found that a large class of models and real networks
although not fractals on all scales 
exhibit fractal properties in boundary shells with $\ell >d$. 
Here we will develop a theory to explain these findings.

\section{Goals of this work}

In this paper, we extend the study of 
network tomography 
describing the shell structure in a randomly connected network with an 
arbitrary degree distribution using generating functions. 
Following Ref. \cite{kalisky}, 
we denote the fraction of nodes at
distance equal to or larger than $\ell$ as
\begin{equation}\label{Eq.rell}
r_\ell \equiv 1-\frac{1}{N} \sum_{m=0}^{\ell-1} B_m,
\end{equation} 
and the nodes at distances equal or larger than $\ell$ as the exterior $E_\ell$ of shell $\ell$. 
Similarly, we define the ``$r-$exterior'', $E_r$, as the $rN$ nodes 
with the largest 
distances from a given root node. To this end, we list all the 
nodes in ascending order of their distances from the root node. 
In this list, the nodes with the same distance are positioned at random. 
The last $rN$ nodes in this list which have the largest distance to the root 
are called the $E_r$.
Notice that $E_r=E_\ell$ if $r=r_\ell$. Introducing $r$ as a continuous 
variable is a new step compared to 
Ref.\cite{kalisky}, which allows us to apply the apparatus of 
generating functions to study network tomography.

The behavior of $B_\ell$ for $\ell<d$ can be approximated by a 
branching process \cite{mnewman}.
In shells with $\ell>d$,
the network will show different topological characteristics compared to 
shells with $\ell<d$. This is due to
the high probability to find high degree nodes (``hubs'') in shells 
with $\ell<d$, 
so there is a depletion of high degree nodes in
the degree distribution in $E_\ell$ with $\ell>d$.
Indeed, the average degree of 
the nodes in shells with $\ell<d$ is greater than the 
average degree in the shells with $\ell>d$ \cite{kalisky,Shao}.

Here, we develop a theory to explain the behavior of the degree distribution 
$P_r(k)$ in $E_r$ and the behavior of the average
degree $\langle k(r) \rangle$ as a function of $r$
in a randomly connected network with a given degree distribution.
Further, we derive analytically $r_{\ell+1}$ as a function of $r_\ell$, 
$r_{\ell+1}=\phi(r_\ell)$, where $\phi$ can be expressed in terms of 
generating functions \cite{Harris1} of 
the degree distribution of the network.
Using these derived analytical expressions, we explain the 
power law distribution
$P(B_\ell)\sim B_\ell^{-2}$ for $\ell\gg d$ found in \cite{Shao}. 
Further, based on our approach, we introduce the network correlation 
function $c(r_\ell)=r_{\ell+1}/\phi(r_\ell)$
to characterize the correlations in the network.
We apply this measure to several model and real-world networks. We find
that the networks fall into two distinct classes: a class of poorly-connected 
networks with $c(r_\ell)>1$, where the virus spreads from a given root
slower than in randomly connected networks with the same degree 
distribution; a class of well-connected networks with $c(r_\ell)>1$, 
in which the virus spreads faster than in a randomly connected network.

In this paper we study RR, ER, SF, WS and BA models, as well as 
several real networks including the Actor collaboration
network (Actor) \cite{barasci}, High Energy Physics 
citations network (HEP) \cite{hep}, the Supreme Court Citation 
network (SCC) \cite{scc} and Autonomous System (AS) Internet 
network (DIMES) \cite{dimes}.
As we will show later, WS, Actor, HEP, and SCC belong to the class
of poorly-connected networks ($c(r_\ell)>1$), 
while BA model and DIMES network 
belong to the class of well-connected networks ($c(r_\ell)<1$).


The paper is organized as follows. In Sec. III, we derive analytically 
the degree distribution and average degree of
nodes in $E_r$ and test our theory on ER and SF networks. 
In Sec. IV, we derive analytically a
deterministic iterative functional form for $r_\ell$. In Sec. V, we 
apply our theory to explain the distribution and average value of 
the number of nodes in shells. In Sec. VI, we
introduce the network correlation function and apply it to 
different networks. Finally, we present summary in Sec. VII.

\section{Degree distribution of nodes in $r$-exterior $E_r$}
\subsection{Generating function for P(k)}
The generating function of a given degree distribution 
$P(k)$ is defined as \cite{Harris1,bingham,braunstein,mnewman},
\begin{equation}\label{Eq.G0}
G_{0}(x)\equiv \sum _{k=0}^{\infty} P(k) x^k.
\end{equation}
It follows from Eq.(\ref{Eq.G0}) that the average degree of the network 
$\langle k \rangle=G^{'}_0(1)$.
Following a randomly chosen link, the probability of reaching a node with
$k$ outgoing links (the degree of the node is $k+1$) is
\begin{equation}\label{Eq.tp}
\tilde P(k)=(k+1)P(k+1)/\sum_{k=0}^{\infty}[(k+1)P(k+1)].
\end{equation}
Notice that
$$
\sum_{k=0}^{\infty}(k+1)x^{k}P(k+1)=\sum_{k=1}^{\infty}kx^{k-1}P(k)=G_0^{'}(x)
$$
and
$$
\sum_{k=0}^{\infty}(k+1)P(k+1)=G_0^{'}(1)=\langle k \rangle,
$$ 
where
$\langle k \rangle$ is the average degree of the network.
The generating function for the
distribution of outgoing links $\tilde P(k)$ is
\begin{equation}\label{Eq.G1}
G_{1}(x)=\sum _{k=0}^{\infty}\tilde P(k) x^{k}=G_{0}^{'}(x)/\langle k\rangle.
\end{equation}
The average number of {\it outgoing} links, also called the branching factor of the network, is
\begin{equation}\label{Eq.tk}
\tilde k\equiv \sum_{k=0}^{\infty}k\tilde P(k)=G_{1}^{'}(1)=G_0^{''}(1)/G_0^{'}(1)=\sum_{k=0}^{\infty}\frac{k(k+1)P(k+1)}{\langle k \rangle}=
\frac{\langle k^2 \rangle-\langle k \rangle}{\langle k \rangle},
\end{equation}

For ER networks, $G_0(x)$ and $G_1(x)$ have the same simple form \cite{mnewman},
\begin{equation}\label{Eq.GER}
G_{0}(x)=G_{1}(x)=e^{\langle k \rangle(x-1)},
\end{equation}
and $\tilde k=\langle k \rangle$.

\subsection{Branching process}

For a randomly connected network, loops can be neglected and the construction
of a network can be approximated by a
branching process \cite{Harris1,bingham,braunstein,mnewman}.
In such a process, an outgoing link, no matter at which shell $\ell$ 
from the root node it starts, has the same
probability $\tilde P(k)$ to reach a node with $k$ 
outgoing links in shell $\ell+1$.
This assumption is very good when $\ell$ is small 
and the preferential selection of the nodes with large degree (hubs) in shell $\ell$ does not
significantly deplete the probability of finding high degree nodes in the further out shells.
However, for $\ell > d$, the probability of finding hubs 
decreases significantly, and so does the average degree $\langle k\rangle$ \cite{kalisky,Shao}.
Another limitation of the branching process as a model of a network is that 
it approximates a network as a tree without loops, 
while in a network loops are likely to form for $\ell>d$.
In order to find an approach that works well for all values of $\ell$, 
we follow Ref. \cite{kalisky} and introduce a 
modified branching process
that takes into account the depletion of large degree nodes and the 
formation of loops.

At the beginning of the process,
we have $N$ separate nodes, and each node has $k$ open links, where $k$ is
a random variable with a distribution $P(k)$. We start to build
the network from a randomly selected node (root).
At each time step, we randomly select an open link from shell $\ell$ of the 
aggregate (root and all nodes already
connected to the root) and connect this open link to another open link.
There are three possible ways to select another open link 
(see Fig.~\ref{fig1}), 
which can belong to
\begin{itemize}
\item[(i)] a free node not yet connecting to the aggregate, 
\item[(ii)] a node in shell $\ell+1$,
\item[(iii)] a node in shell $\ell$. 
\end{itemize}
When all the open links from shell $\ell$ are connected, we will then
select an open link from shell $\ell+1$. 
By doing this, the aggregate keeps growing shell 
after shell until all open links are connected. In cases (ii) and (iii), there
are chances to create parallel links (two links connecting a
pair of nodes) and circular links (one link with two ends connected to the same node).
For a large network with a finite branching factor $\tilde k$, 
such events occur with negligible probability.

We denote by $r \equiv r(t)$ \cite{note1} the fraction of distant nodes not
connected to the aggregate at step $t$. These nodes constitute the 
r-exterior $E_r$.
At the beginning of the growth process, before
we start to build the first shell, $r(0)=(N-1)/N\approx 1$.
At the end of the growth process, $r(t)=r_{\infty}$, where $r_{\infty}$ is the fraction of 
nodes that are not 
connected to the aggregate when the building process is finished, i. e., when
all open links in the aggregate are used. 
The process described above
simulates a randomly connected network, which is a good approximation for 
many model and real-world networks.

\subsection{Degree distribution and average degree of nodes in the r-exterior $E_r$}

Let $A_{r}(k)$ be the number of nodes with
degree $k$ in the r-exterior $E_r$ at time $t$. 
The probability to have a node with degree $k$ in $E_r$ is given by \cite{footnote1}
\begin{equation} \label{Eq.Pk}
P_{r}(k)=\frac{A_{r}(k)}{rN}.
 \end{equation}
When we connect an open link from the aggregate to a free node (case (i)),
$A_{r}(k)$ changes as
\begin{equation}\label{Eq.Afk}
A_{r-\frac{1}{N}}(k)=A_{r}(k)-\frac{P_{r}(k)k}{\langle k(r) \rangle},
\end{equation}
where $\langle k(r) \rangle=\sum P_{r}(k)k$ is the average degree of nodes in $E_r$.
In the limit of $N\to \infty$, Eq.(\ref{Eq.Afk}) can be presented as 
the derivative
of $A_{r}(k)$ with respect to $r$ 
\begin{equation}\label{Eq.dAfk}
\frac{dA_{r}(k)}{dr} \approx N[A_{r}(k)-A_{r-\frac{1}{N}}(k)]=N\frac{P_{r}(k)k}{\langle k(r) \rangle}.
\end{equation}
Differentiating Eq.(\ref{Eq.Pk}) with respect to $r$, and using Eq.(\ref{Eq.dAfk}),
we obtain
\begin{equation}\label{Eq.evolf}
-r \frac{dP_{r}(k)}{dr}=P_{r}(k)-\frac{kP_{r}(k)}{\langle k(r) \rangle},
\end{equation}
which is rigorous for $N\to \infty$.
Substituting
\begin{equation}\label{Eq.uG0}
f\equiv G_{0}^{-1}(r)
\end{equation}
in Eq. (\ref{Eq.evolf}), we find by direct differentiation that
\begin{equation} \label{Eq.Pku}
P_{f}(k)=P_{1}(k) \frac{f^{k}}{G_{0}(f)},
\end{equation}
and
\begin{equation}\label{Eq.ku}
\langle k(f) \rangle=\frac{fG_{0}^{'}(f)}{G_{0}(f)},
\end{equation}
is the solution satisfying Eq. (\ref{Eq.evolf}). Notice 
that $P_{1}(k) \equiv P(k)$.

Eq. (\ref{Eq.Pku}) and Eq. (\ref{Eq.ku}) are respectively the degree 
distribution and the average degree in $E_r$,  
as functions of $f$. Once we know the explicit functional form for $G_0(x)$,
we can invert $G_0(x)$ to find $f=G_0^{-1}(r)$ and find
analytically both $P_r(k)$ and $\langle k(r) \rangle$:
\begin{equation} \label{Eq.Pku2}
P_{r}(k)=P(k)\frac{[G_0^{-1}(r)]^k}{r},
\end{equation}
\begin{equation}\label{Eq.ku2}
\langle k(r) \rangle=\frac{G_0^{-1}(r)G_0^{'}( G_0^{-1}(r))}{r}.
\end{equation}
In a network with minimum degree $k_{\rm min}\geq 2$, we find by Taylor
expansion that
\begin{equation}\label{Eq.kuapp}
\langle k(r) \rangle= k_{\rm min}+\frac{P(k_{\rm min}+1)}{P(k_{\rm min})^{1+\alpha}} r^{\alpha}+O(r^{2\alpha}),
\end{equation}
where $\alpha\equiv 1/k_{\rm min}$.

For ER networks, using Eq. (\ref{Eq.GER}) and
Eq. (\ref{Eq.ku}), we find 
\begin{equation}\label{Eq.kf}
\langle k(r) \rangle=\ln r +\langle k \rangle.
\end{equation}
For $0<r\leq 1$, Eq. (\ref{Eq.Pku}) can be rewritten as
\begin{equation} \label{Eq.Pkf}
P_{r}(k) = P(k) \frac{(\ln r /\langle k \rangle +1)^{k}}{r}= e^{- \langle k(r) \rangle}  \frac{\langle k(r) \rangle ^k}{ k !},
\end{equation}
which implies that the degree distribution in the distant nodes remains a 
Poisson distribution but with a smaller average degree $\langle k(r) \rangle$.

Next, we test our theory numerically for ER networks with $N=10^6$ nodes and
different values of $\langle k \rangle$.
To obtain $P_r (k)$, we start from a randomly chosen root node,
and find the nodes in $E_r$ and
their degree distribution $P_r(k)$.
This process is repeated many times for different roots and different
realizations.
The results are shown in Fig.~\ref{fig2}a. 
The symbols are the
simulation results of the degree 
distribution in $E_r$ for $r=1$, 0.5 and 0.05. 
The analytical results (full lines) are computed using Eq. (\ref{Eq.Pkf}).
As can be seen, the theory agrees very well with the simulation 
results for both $r=$0.5
and 0.05. We compared our theory with the simulations also 
for other values of $r$ and
$\langle k \rangle$ and the agreement is also excellent.

For SF networks, $G_0(x)$ and $G_1(x)$ cannot be expressed as
 elementary functions \cite{mnewman}.
But for a given $P(k)$,
they can be written as power series 
of $x$ and one can compute the expressions 
in Eq.(\ref{Eq.Pku}) and Eq.(\ref{Eq.ku}) numerically.
In order to reduce the systematic errors caused by estimating $P(k)$, 
we write $G_0(x)$ and $G_1(x)$ 
based on the $P(k)$ obtained from the simulation results instead 
of using its theoretical form.

We built SF networks using
the Molloy-Reed algorithm \cite{MR}.
In Fig.~\ref{fig2}b, the symbols represent the simulation results for $P_r(k)$ obtained
for $E_r$ of SF network with $\lambda=3.5$ and  
$r=1$, 0.5 and 0.1.
The lines are the numerical results calculated from Eq.(\ref{Eq.Pku}). 
Good agreement between
the simulation results and the theoretical predictions can be seen in Fig.~\ref{fig2}b.
Other values of $r$ and $\lambda$ have also been tested with good 
agreement.

In Fig.~\ref{fig3}a, we show the average degree $\langle k(r) \rangle$ in
$E_r$ as a function of $r$ for ER networks with different
values of $\langle k \rangle$. 
Lines representing Eq.(\ref{Eq.kf})
agree very well with the numerical results (symbols) even
for very small $r$.
We note that Fig.~\ref{fig3}a shows different value of lower 
limit cutoff $r_{\infty}$ 
for $r$, when
$\langle k(r) \rangle$ is very small. 
As mentioned before, $r_{\infty}$ is the fraction of
nodes which are not connected to the aggregate at the end of the process. 
In the next section, we will present an equation for $r_{\infty}$.

In Fig.~\ref{fig3}b, we present the numerical 
results of Eq.(\ref{Eq.ku}) for SF 
networks with different values of $\lambda$.
For a given $E_r$, $\langle k(r) \rangle$ is computed
from the simulated network and the results are averaged
over many realizations. Good agreement between
the theory (lines) and the simulation results (symbols) can be seen.

\section{iterative functional form of $r_\ell$, the fraction of nodes outside shell $\ell$}
In this section, we study the growth of the aggregate itself.
Let $L(t)$ be the number of open links belonging to the full aggregate at
step $t$, and $\Lambda(t) \equiv L(t)/N$.
The number of open links belonging to shell $\ell$ of the 
aggregate is
defined as $L_{\ell}(t)$ and $\Lambda_{\ell}(t) \equiv L_{\ell}(t)/N$.
After we finish building shell $\ell$ and just before we start to  
build shell $\ell+1$, all the open links in the aggregate belong to 
nodes in shell $\ell$,
so $t=t_\ell$, we have $\Lambda_{\ell}(t)=\Lambda(t)$ \cite{footnote2}.
In the process of building shell $\ell+1$, $\Lambda_{\ell}(t)$
decreases to $0$. 

Next we show that both $\Lambda(t)$ and $\Lambda_{\ell}(t)$ can  
be expressed as functions of $r$. 
In analogy with Eq.(\ref{Eq.tk}), we define the branching factor of nodes in the 
r-exterior $E_r$ as
\begin{equation}\label{Eq.tkf}
\tilde{k}(r)=\frac{\langle k^2(r)\rangle-\langle k(r)\rangle}{\langle k(r)\rangle}=\frac{\sum _{k=0}^{\infty}k^{2}P_{r}(k)}{\langle k(r)\rangle}-1.
\end{equation}
Using Eq.(\ref{Eq.tkf}) and Eq.(\ref{Eq.ku}), 
$\tilde{k}(r)$ can be rewritten as a
function of $f$ as
\begin{equation}\label{Eq.tku}
\tilde{k}(f)=\frac{fG_{0}^{''}(f)}{G_{0}^{'}(f)}.
\end{equation}

Appendix \ref{A1} shows that $\Lambda(r)$ 
and $\Lambda_{\ell}(r)$ obey differential equations  
\begin{eqnarray}
    \frac{d\Lambda(r)}{dr}&=-\tilde{k}(r)+1+\frac{2\Lambda(r)}{r\langle k(r)\rangle} \label{Eq.dequations1} \\
   \frac{d\Lambda_{\ell}(r)}{dr}&=1+\frac{\Lambda(r)}{r\langle k(r) \rangle}+\frac{\Lambda_{\ell}(r)}{r\langle k(r) \rangle}  \label{Eq.dequations2}
\end{eqnarray}

Eq.(\ref{Eq.dequations1}) and Eq.(\ref{Eq.dequations2})
govern the growth of the aggregate.
To solve them, we make the same substitution
$f=G_{0}^{-1}(r)$ (Eq.[\ref{Eq.uG0}]) as before.
The general form of the solution for Eq.(\ref{Eq.dequations1}) is
\begin{equation}\label{Eq.sdlfp}
\Lambda(f)=-G_0^{'}(f)f+C_{1}f^{2},
\end{equation}
 where $C_1$ is a constant. At time t=0, $r=f=1$, and $\Lambda(1)=0$.
 With this initial condition,
we obtain $C_1=G_{0}^{'}(1)=\langle k \rangle$.
Using Eq.(\ref{Eq.sdlfp}), the general solution of 
Eq.(\ref{Eq.dequations2}) is  
\begin{equation}\label{Eq.sdllfp}
\Lambda_{\ell}(f)=G_{0}^{'}(1)f^{2}+C_2 f,
\end{equation}
where $C_2$ is a constant. When $r=r_{\ell}$, the building of shell $\ell$ is finished. At that
time, all the open links of the aggregate
 belong to shell $\ell$, $\Lambda(r)\mid _{r=r_{\ell}}=\Lambda_{\ell}(r)\mid _{r=r_{\ell}}$.
If we denote $f_{\ell}\equiv G_0^{-1}(r_{\ell})$, $C_2=-G_0^{'}(f_{\ell})$.
Thus, the solutions of the differential equations
 Eqs.(\ref{Eq.dequations1}) and (\ref{Eq.dequations2}) are
\begin{eqnarray}
   \Lambda(f)&=G_{0}^{'}(1) f^2-G_{0}^{'}(f)f \label{Eq.sol1}\\
   \Lambda_{\ell}(f)&=G_{0}^{'}(1) f^{2}-G_{0}^{'}(f_{\ell})f  \label{Eq.sol2}
\end{eqnarray}

When all open links in the aggregate are used, $\Lambda=0$, 
the corresponding $f=f_\infty$ gives the fraction of nodes 
$r_\infty=G_0(f_\infty)$
which do not belong 
to the aggregate when the building process is finished.
The value of $f_\infty$ must satisfy Eq.(\ref{Eq.sol1}) with $\Lambda(f_\infty)=0$
\begin{equation}\label{Eq.uinfty}
f_{\infty}=G_{0}^{'}(f_{\infty})/G_{0}^{'}(1)\equiv G_{1}(f_{\infty}),
\end{equation}
and from Eq.(\ref{Eq.uG0})
\begin{equation}\label{Eq.efin}
r_{\infty}=G_0(f_{\infty}). 
\end{equation}
Eqs. (\ref{Eq.uinfty}) and (\ref{Eq.efin}) imply that there exist a certain 
fraction of distant links and nodes not connected to the aggregate when the building process 
is finished.
These results are consistent with previous work \cite{braunstein}.
The numerical solution for Eq.(\ref{Eq.uinfty}) 
 is discussed in Sec. IV (A) and Appendix B.

When $\Lambda_{\ell}(f)=0$, the construction of shell $\ell+1$ is completed, 
$r=r_{\ell+1}$ and $f=f_{\ell+1}$. Then
from Eq.(\ref{Eq.sol2}), we obtain
\begin{equation}\label{Eq.ul1}
f_{\ell+1}=G_{0}^{'}(f_{\ell})/G_{0}^{'}(1)=G_{1}(f_{\ell}),
\end{equation}
which leads to a deterministic iterative functional form for $r_{\ell}$
\begin{equation}\label{Eq.fl1}
 r_{\ell+1}=G_{0}(f_{\ell+1})=G_{0}(G_{1}(G_{0}^{-1}(r_{\ell}))\equiv 
\phi(r_{\ell}).
\end{equation}
Eq. (\ref{Eq.fl1}) allows us to make a deterministic prediction of $r_{\ell+1}$
once we know $r_{\ell-1}$.

This result is different from a similar well-known result \cite{Harris1} 
based on the 
physical meaning of the generating function $G_0(r)$, which gives a fraction 
of nodes in the set B not directly connected to a randomly selected fraction $1-r$ of 
set A. The difference with Eq.(\ref{Eq.fl1}) is that set A is selected 
not by constructing shells around a root but randomly. Moreover, set B 
may even overlap with set A.

To test our theory,
we use RR networks, where $P(k)=\delta(k-\psi)$, $G_0(x)=x^{\psi}$ 
and $G_1(x)=x^{\psi-1}$,
then Eq.(\ref{Eq.fl1}) reduces to
\begin{equation} \label{Eq.ffrl}
r_{\ell+1}=r_{\ell}^{\psi-1},
\end{equation}
which is shown as lines in Fig.~\ref{fig4}a. The symbols in 
Fig.~\ref{fig4}a are the
simulation results for RR networks with different values of $\psi$. 
To obtain the simulation
results, at each realization a
random root is chosen and a full set of $r_{\ell}$ is computed. The results 
obtained for many realizations are plotted as a scatter plot.
Due to the homogeneity of RR network, 
$r_{\ell}$ can only take on discrete values.
The agreement between the simulation results 
and Eq.(\ref{Eq.ffrl}) is excellent, and 
the scattering almost cannot be observed \cite{scatter}.

For ER networks, Eq.(\ref{Eq.fl1}) yields
\begin{equation} \label{Eq.ffa}
r_{\ell+1}=e^{\langle k \rangle(r_{\ell}-1)},
\end{equation}
which is valid for all $\ell>1$. We test Eq.(\ref{Eq.ffa}) for ER network with different
values of $\langle k \rangle$ and the results are shown in Fig.~\ref{fig4}b. 
The agreement between the theoretical predictions (lines) and the the 
simulation results is excellent.

For SF networks, Eq.(\ref{Eq.fl1}) can be solved numerically using the 
values of $P(k)$ from the generated SF network. The lines shown
in Fig.~\ref{fig4}c represent the numerical solutions of 
the theory [Eq.(\ref{Eq.fl1})]. The symbols
are the simulation results for the generated SF networks. 
For $\lambda\geq 2.5$, a good agreement between
theory and simulation results can be seen. 
Note that for the very small value of $\lambda=2.2$,
the simulation results deviate slightly from the 
theory due to high probability of creating {\it parallel and circular 
links} (PCL) in the hubs of the randomly connected network \cite{boguna} 
(created by case (ii) and (iii) in Fig.~\ref{fig1}).
We test Eq.(\ref{Eq.fl1}) for a SF network 
of $\lambda=2.2$ allowing PCL during its construction. 
The results are shown in Fig.~\ref{fig4}d as a log-linear plot. 
The agreement between the theory and the simulation results for a SF network
with $\lambda=2.2$ in presence of PCL is very good.
This shows that SF networks built by the Molloy-Reed algorithm 
without PCL deviate
from randomly connected networks for very small values of $\lambda$. 
We will further discuss
this deviation in Sec. V B.

\section{applications}

\subsection{Derivation of the power-law distribution of $B_\ell$
for $\ell\gg d$}

Recently, a broad power-law distribution of the number of nodes at
 shell $\ell$
($\ell\gg d$), $B_{\ell}$, has been reported \cite{Shao}. This power-law
distribution exists in many model and real networks and is characterized by 
a universal form $P(B_{\ell})\sim B_{\ell}^{-2}$ (see Fig.~\ref{figab}). 
Using Eq.(\ref{Eq.fl1}),
we will prove this relation and explain the origin of this universal 
power-law distribution.

For the purpose of clarity, we use $m$ instead of $\ell$ for 
shells with $\ell<d$, and $n$ instead of $\ell$ 
for $\ell>d$. 
For the entire range of shell indices, $\ell$ will be used.

For infinitely large networks, 
we can neglect loops for $\ell<d$ and approximate
the forming of a network as a
branching process \cite{Harris1,bingham,braunstein,mnewman}.
It has been reported \cite{mnewman, bingham} that for shell $m$ 
(with $m\ll d$),
the generating function for the number of nodes, $B_{m}$,
in the shell $m$ is
\begin{equation}\label{tildegm}
\tilde{G}_{m}(x)=G_{0}(G_{1}(...(G_{1}(x))))=G_{0}(G_{1}^{m-1}(x)),
\end{equation}
where $G_{1}(G_{1}(...)) \equiv G_{1}^{m-1}(x)$ is the result of
applying $G_{1}(x)$, $m-1$ times and 
$P(B_m)$ is the coefficient of $x^{B_{m}}$ 
in the Taylor expansion of $\tilde{G}_m(x)$ around $x=0$.
The average number of nodes in shell $m$ is $\tilde k^m$ 
\cite{mnewman}.
It is possible to show that $G_{1}^{m}(x)$
converges to a function of the form
$\Phi((1-x)\tilde{k}^m)$ for large $m$ \cite{bingham}, where $\Phi(x)$ satisfies the
Poincar\'{e} functional relation
\begin{equation} \label{Eq.gphi}
G_{1}(\Phi(y))=\Phi(y \tilde{k}),
\end{equation}
where $y\equiv 1-x$. The functional form of $\Phi(y)$ can be uniquely determined from
Eq.(\ref{Eq.gphi}).

It is known that $\Phi(y)$ has an asymptotic functional form,
$\Phi(y)=f_\infty+ay^{-\delta}+o(y^{-\delta})$, where $a$ is a constant \cite{bingham}.
Expanding both sides of Eq.(\ref{Eq.gphi}), we obtain
\begin{equation}\label{gexpend}
G_{1}(f_{\infty})+G_{1}^{'}(f_{\infty})ay^{-\delta}=f_{\infty}+a\tilde{k}^{-\delta}y^{-\delta}+o(y^{-\delta}).
\end{equation}
Since $G_{1}(f_{\infty})=f_{\infty}$, we find
\begin{equation}\label{Eq.delta}
\delta=-\ln G_{1}'(f_\infty)/\ln\tilde{k}.
\end{equation}
The numerical solution of $G_{1}(f_{\infty})=f_{\infty}$ depends on different scenarios 
(see Appendix \ref{B1}) as
\begin{equation} \label{Eq.f00}
f_{\infty} \left\{%
\begin{array}{ll}
    >0, & \hbox{for $P(k=1)\neq 0$;} \\
    =0, & \hbox{for $P(k=1)=0$.} \\
   \end{array}%
\right.
\end{equation}
The solution for $\delta$ is (see Appendix \ref{B1})
$$
\delta \left\{%
\begin{array}{ll}
    >0, & \hbox{for $P(k=1)\neq 0$ and $P(k=2)\neq 0$;} \\
    =\infty, & \hbox{for $P(k=1)= 0$ and $P(k=2)= 0$.} \\

\end{array}%
\right.
$$

Applying Tauberian-like theorems \cite{bingham,dubuc} to $\Phi(y)$, which has a
power-law behavior for $y\to \infty$, 
the Taylor expansion coefficient of $\tilde{G}_{m}(x)$,
it has been found \cite{dubuc} that $P(B_{m})$ behaves
as $B_{m}^{\mu}$ with an exponential cutoff at $B_{m}^{*}\sim \tilde{k}^{m}$
and some quasi-periodic modulations with period 1 as a function of 
$\log_{\tilde k}B_{m}$ 
\cite{bingham,dubuc}, where
$$
\mu= \left\{%
\begin{array}{ll}
    \delta-1, & \hbox{for $P(k=1)\neq 0$ ;} \\
    2\delta-1, & \hbox{for $P(k=1)=0$ and $P(k=2)\neq 0$ ;} \\
    \infty, & \hbox{for $P(k=1)=0$ and $P(k=2)= 0$ .} \\
\end{array}%
\right.
$$
Thus, the probability distribution of the number of nodes
in the shell $m$ has a power law tail for
small values of $B_{m}$ \cite{Shao},
\begin{equation}\label{Eq.pbm}
P(B_{m})\sim B_{m}^{\mu},
\end{equation}
if $P(k=1)+P(k=2)>0$. 

The above considerations are correct only for $m\ll d$, where the depletion of nodes with
large degree is insignificant.
For $\ell>d$, we must consider the changing of $P_r(k)$.

Using Eq.(\ref{Eq.fl1}) for the whole range of $\ell$,
we can write the relation between $r_n$ for $n>d$ and
$r_m$ for $m\ll d$ as
\begin{equation}\label{Eq.fnm}
r_{n}=G_{0}(G_{1}(G_{0}^{-1}(G_{0}(G_{1}(G_{0}^{-1}...(r_{m})...)=G_{0}(G_{1}^{n-m}(G_{0}^{-1}(r_{m})))=G_{0}(G_{1}^{n-m}(f_m)).
\end{equation}
Applying the same considerations as for $B_m$, we obtain,
\begin{equation}\label{Eq.gfm}
G_{1}^{n-m}(f_m) = f_{\infty}+a\tilde{k}^{-\delta(n-m)}(1-f_m)^{-\delta}.
\end{equation}
Using 
\begin{equation}\label{Eq.1mfm}
1-f_m=1-G_{0}^{-1}(r_{m})=1-G_{0}^{-1}(1-(1-r_{m})), 
\end{equation}
we can write a 
Taylor expansion for $z\equiv 1-r_m$ as 
\begin{equation}\label{Eq.1g0}
1-f_m=1-G_{0}^{-1}(1+z)\approx 1-[1-z(G_0^{-1})^{'}(1)]
=z/\langle k \rangle.
\end{equation}
Thus, we obtain
\begin{equation}\label{Eq.gfm2}
G_{1}^{n-m}(f_m) \approx f_{\infty}+a[\frac{\tilde{k}^{n-m}}{\langle k \rangle}(1-r_m)]^{-\delta}.
\end{equation}
Applying $G_0$ on both sides of Eq.(\ref{Eq.gfm2}) and using Taylor expansion,
we obtain
\begin{equation}\label{Eq.fnm2}
r_n=G_0(f_{\infty})+G_0^{'}(f_\infty) \zeta+ \frac{G_0^{''}(f_\infty)}{2} \zeta^2...,
\end{equation}
where $\zeta\equiv a(\tilde{k}^{n-m}/ \langle k \rangle)^{-\delta}(1-r_m)^{-\delta}$.
If $P(k=1)\neq 0$, as discussed in Eq.(\ref{Eq.f00}) and Appendix B, 
$f_\infty$ is non-zero,
$G_0^{'}(f_\infty)=\langle k \rangle G_1(f_\infty)=\langle k \rangle f_\infty$
is also non-zero, thus we can ignore the $\zeta^2$ term and keep the
leading non-zero term $\zeta$. 
If $P(k=1)=0$ and $P(k=2)\neq 0$,
both $G_1(f_\infty)$ and $f_\infty$ are zero,
$G_0^{''}(f_\infty)=\langle k \rangle G_1^{'}(0)=2P(k=2)\neq 0$ and then
$\frac{G_0^{''}(f_\infty)}{2} \zeta^2$
is the leading non-zero term.
Thus,
\begin{eqnarray}
  r_n -r_\infty \approx af_\infty \frac{\tilde{k}^{\delta(n-m)}}{\langle k \rangle^
     {-\delta-1}}(1-r_m)^{-\delta}\approx(1-r_m)^{-\mu-1} , & P(k=1)\neq 0 \label{Eq.40}\\
   r_n -r_\infty \approx P(2)a^2\left[\frac{\tilde{k}^{n-m}(1-r_m)}{\langle k \rangle}\right]^{-2\delta}\approx (1-r_m)^{-\mu-1}, & P(k=1)=0, P(k=2)\neq 0 \label{Eq.41}
\end{eqnarray}

Since $B_\ell$ increases exponentially with $\ell$ for $\ell<d$ 
and decreases even faster than
exponentially for $\ell>d$ \cite{mnewman}, we can make approximations
$r_n \sim B_n/N$ and $1-r_m \sim B_m/N$ for $n\gg d$ and $m\ll d$ respectively. Using
$P(B_n)dB_n=P(B_m)dB_m$ and Eqs.(\ref{Eq.pbm}), (\ref{Eq.40}) and (\ref{Eq.41}), we obtain
 \begin{equation}\label{Eq.pbn}
P(B_n)\sim B_n^{-1-\mu /(\mu+1)-1/(\mu+1)}=B_{n}^{-2},
\end{equation}
which is valid for $n\gg d$.

The power-law distribution shown by Eq.(\ref{Eq.pbn}) 
indicates that fractal features exist
at the boundaries of almost all networks. Further studies of 
these fractal features are represented in Ref. \cite{Shao}.

\subsection{Average number of nodes in shell $\ell$, $\langle B_\ell \rangle$}

The number of nodes in shell $\ell$ can be expressed
 as a function of $r_\ell$ as
\begin{equation}\label{Eq.bll}
B_\ell=N(r_\ell-r_{\ell+1}).
\end{equation}
From Eq.(\ref{Eq.fl1}) and Eq.(\ref{Eq.bll}),
with initial condition $r=r_m$,
one can calculate $B_\ell$ for all $\ell\geq m$
and find $\langle B_\ell \rangle$
for $\ell\geq m$ using $P(B_\ell)$.

However, when we study a finite network, the effect of the first few shells needs to be
considered. Take a RR network as an example. From simulation data it is clear that
$B_0=1$, $B_1=\psi$ and $B_2=\psi(\psi-1)$,
and correspondingly $r_1=1-1/N$, $r_2=1-(\psi+1)/N$ and $r_3=1-(1+\psi+\psi (\psi-1))/N=
1-(\psi^2+1)/N$.
If we apply Eq.(\ref{Eq.ffrl}) on $r_1$ and $r_2$, 
the calculated $r_2=1-(\psi-1)/N$ and
$r_3=1-(\psi^2-1)/N$ deviate from 
the simulated results of $r_2$ and  $r_3$ 
by a constant value of $2/N$. For $N\to \infty$ and large $\ell$,
this deviation is negligible. However for a finite system and
small $\ell$, we have to consider this
term. To cancel this constant deviation, we modify Eq.(\ref{Eq.ffrl}) as
\begin{equation}\label{Eq.ffrlm}
r_{\ell+1}=r_{\ell}^{\psi-1}-2/N.
\end{equation}
Using Eq. (\ref{Eq.ffrlm}), starting from $r_1$, we can calculate $r_\ell$ and $B_\ell$ for any $\ell>1$.
For RR network, due to 
the homogeneity of the degree, the distribution of $B_\ell$
is a delta function, thus $\langle B_\ell \rangle=B_\ell$.
In Fig.~\ref{fig7}, we show the theoretical predictions of 
$\langle B_\ell\rangle$ (full lines)
together with the simulation results (symbols) for different values of $\psi$.
The simulation results
are the average over different realizations. The agreement between
the theory and the simulation results is excellent.

For networks with varying degree (like ER and SF), $\langle B_\ell\rangle $ 
cannot be directly calculated from
our theory. The reason is that for these networks, the
modification needed on Eq.(\ref{Eq.fl1}) is not a constant but fluctuates with a 
magnitude of the order of $1/N$. Further, because
$\langle \phi(r_\ell)\rangle \neq  \phi(\langle r_\ell\rangle)$, 
we cannot replace $B_\ell$ with $\langle B_\ell\rangle$ as we did for RR.
As we see in
Fig.~\ref{fig4}, Eq.(\ref{Eq.fl1}) works well also 
for varying degree networks
in predicting $r_{\ell+1}$ once $r_{\ell}$ is known. It also works well in predicting $B_{\ell+\Delta}$
($\Delta=1,2,3...$) given a shell with big enough $B_\ell$ ($\approx 10^4$). It can 
reproduce the behavior of successive shells with $99\%$ accuracy. However, when $B_{\ell+\Delta}$ become
small ($<10^4$), the error is relatively large.

\section{The Network correlation function $c(r)$}

In this section we will compare various models and real-world networks with the 
randomly connected networks with same degree distributions and 
introduce a new network characteristic, the network correlation 
function $c(r)$ analogous
to the density correlation function in statistical mechanics \cite{hansen}. 
For a randomly
connected network, $c(r)=1$, as for the density correlation function 
in the ideal gas, while for the non-random networks the deviation of $c(r)$ from unity characterizes their 
correlations on different distances from the root. 
    
\subsection{SF networks with $\lambda \leq 3$}

Our theory crucially depends on the existence of the branching factor $\tilde{k}$. 
So we can expect significant deviations from our theory in the behavior of the SF networks 
with $\lambda\leq 3$, for which $\tilde{k}$ diverges
for $N\to\infty$. However, for a fixed $N$, the degree distribution is truncated
by the natural cutoff $k_{\rm max}\sim N^{1/(\lambda-1)}$, so that $\tilde{k}$ 
still exists. Hence, we 
hypothesize that our theory remains valid even for $\lambda<3$ for 
randomly connected networks (with PCL) (see Fig.~\ref{fig6a}).
Another problem is that our algorithm 
of constructing randomly connected networks
leads to formation of PCL. 
The PCL is typically forbidden in the construction algorithms of the network characterizing
complex systems. 
In order to
construct a network without PCL, one imposes significant
correlations in network structure of a dissortative nature with greater probability of hubs 
to be connected to small degree nodes than in a randomly connected
network \cite{boguna}. Thus, we can predict that         
SF networks with $\lambda\leq 3$ which do not include PCL 
must significantly deviate from the prediction of our theory. 

In order
to characterize this deviation we define a correlation function 
\begin{equation}\label{Eq.rg}
c(r_\ell)\equiv r_{\ell+1}/\phi(r_\ell), 
\end{equation}
where $r_{\ell+1}$ and $r_\ell$ characterize two successive shells of a network 
under investigation while $\phi(r_\ell)$ is the prediction (Eq.(\ref{Eq.fl1})) 
of $r_{\ell+1}$ based on our theory for a randomly connected network. Accordingly, 
we compute $c(r_\ell)$ for several networks
with $N=10^6$ nodes with $\lambda=2.5$ and $\lambda=2.2$, for the randomly 
connected case and for the case in which PCL are not allowed. We find in Fig.~\ref{fig6a} 
that for randomly connected networks $c(r_\ell)$ is always
close to 1 with the expected random deviations for 
$r_\ell\to 0$ and $r_\ell\to 1$ caused by random fluctuations in the small first
($r_\ell\to 1$) and last ($r_\ell\to 0$) shells. In contrast, $c(r_\ell)$ is
uniformly smaller than 1 for the networks without PCL.
For $\lambda=2.5$ the deviations are small because the typical number of PCL
that would randomly form still constitute a negligible fraction of links. 
For $\lambda=2.2$ the deviations are significant because
in this case the chance of formation of PCL is much higher. In both cases, the deviation 
are increasing with the maximal degree of the network, which can randomly fluctuate around 
its average value $k_{\rm min}N^{1/(\lambda-1)}\Gamma[(\lambda-2)/(\lambda-1)]$ \cite{cohen2000}. 
The value of $c<1$ for 
these networks indicates the fact that due to the absence of PCL more nodes are attached to 
the next shells compared to randomly connected networks. Accordingly, for such networks, 
the fraction of nodes not included into
shell $\ell+1$ is smaller than that in randomly connected networks. 
Thus, SF networks for $\lambda<3$ are dissortative, which means that the degree of
a node is anti-correlated with the average degree of its neighbors. Moreover,
these anti-correlations are barely visible for $\lambda >2.5$ and increase with 
the decrease of $\lambda$. Therefore $c(r_\ell)<1$ can be associated with the network dissortativeness.

\subsection{Global measurement of correlations}

The network building process described in Sec. II corresponds to a randomly 
connected network for a given
$P(k)$. However, real-world and model networks do not always follow the 
behavior described by our theory. The correlation function $c(x)$ 
constructed in the previous section [Eq. (\ref{Eq.fl1})] can be used to 
detect non-randomness in the network connections. 

For a given degree distribution, we define poorly-connected networks as
those in which $c(r_\ell)>1$. Conversely, we define well-connected
networks as those in which $c(r_\ell)<1$. The motivation for this 
definition is that if $c(r_\ell)>1$, it means that the number of nodes in
shell $\ell$, 
$B_{\ell}=N(r_{\ell}-r_{\ell+1})= N[r_{\ell}- c(r_\ell)\phi(r_\ell)]$, 
is smaller than $N[r_{\ell}-\phi(r_\ell)]$, the value expected for a randomly connected network with the same degree distribution. 
Therefore in a poorly-connected network information or virus spreads slower 
than in a randomly connected network in accordance with the meaning of the 
term poorly-connected.
Conversely, in well-connected networks information spreads faster than in 
randomly connected network with the same degree distribution. Poorly-connected
networks usually contain cliques of fully connected nodes. 
In a clique, the
majority of links connect back to the already connected nodes in shell $\ell$. So the new
shell $\ell+1$ grows slower than for a randomly connected network with the 
same degree distribution. 

As an example, we analyze the WS model 
characterized by high clustering. In this case the number of links which 
can be used to build the next 
shell of neighbors is much smaller than in a randomly connected network 
with the same 
degree distribution.  Thus we can expect $c>1$ in particular for a 
small fraction $\beta$ of 
rewired links (see Fig.~\ref{fig6}a). 
Further, we find that the networks characterizing 
human collaborations are usually poorly connected (see Fig.~\ref{fig6}b).
A typical example of such a network is the actor network, where a link between 
two actors indicates that they play in the same movie at least once.    
So all the actors played in the same movie form a fully connected 
subset of the network (``clique''). As a result, the majority of their links 
are not used to
attract new actors but circle back to the previously acquainted actors. 
The same is correct for the Supreme Court Citation network (SCC) 
and High Energy Physics 
citations (HEP) networks in Fig.~\ref{fig6}b. 
Actor, HEP, and SCC networks all contain a  
large amount of highly inter-connected cliques. 
As we see these cliques manifests themselves in $c>1$.  
In contrast the DIMES network 
\cite{dimes}, is designed to be well-connected and as 
a result it has $c<1$.
   
Another example of a well-connected network is the BA model, 
in which $c(r_\ell)$ linearly goes to zero for $r_\ell\to 0$ [
Fig. \ref{fig6}(c)]. In the BA model a new node, which has exactly $k_{\rm min}$
open links, randomly attaches its links to the previously existing nodes with
probabilities proportional to their current degrees. 
(PCL are forbidden.) 
One can see that for $k_{\rm min}\geq 2$, $c(r_{\ell})<1$
for all $r_\ell$ except in a small vicinity of $r_\ell= 1$. This fact
is associated with the dissortative nature of the BA model, in which small degree nodes
that are created at the late stages of the network construction 
are connected with very high probability to the hubs that are created at the 
early stages. Thus as soon as the hubs are reached during shell construction, 
the rest of the nodes can be reached much faster than in a randomly connected 
network. 

The small region of $c(r_{\ell})>1$ for $r_{\ell}\to 1$ can be associated with 
the fact that the hubs which are created at the early stages of the BA network 
construction, are not necessarily directly connected to each other as it 
would be in randomly connected networks. Thus initial shells of the BA model 
corresponding to large $r_\ell$ grow slower than they would grow in the 
randomly connected network. The effect is especially strong for
$k_{\min}=1$ in which the BA network is a tree, and the distance between 
certain hubs can be quite large. Thus BA with $k_{\rm min}=1$ gives an example 
of a network with poor connectivity between the hubs (large $r_\ell\to 1$) 
and good connectivity among the low degree nodes ($r_\ell\to 0$) which are 
directly connected to the hubs. In a network in which long connected chains
of low degree nodes are abundant, we will have poor 
connectivity ($c(r_\ell)>1$) for $r_\ell \to 0$. In general, the behavior 
of $c(r_\ell)$ for $r_\ell\to 1$ characterizes the connectivity 
among the hubs, while the behavior of $c(r_\ell)$ for $r_\ell\to 0$ 
characterizes the connectivity among the low-degree nodes.

\section{summary}
In this paper, we derive new analytical relations describing shell 
properties of a randomly connected network. In particular, we expand the 
results of Ref. \cite{kalisky} on the network tomography using the 
apparatus of the generating functions.  We find how the degree distribution 
is depleted as we approach the boundaries of the network 
which consist of the $r$-fraction of nodes which are most distant from a root node. 
We find an explicit analytical expression for the degree distribution 
as a function of $r$ [Eqs. (\ref{Eq.Pku}) and (\ref{Eq.ku})]. We also derive 
an explicit analytical relation between the values of $r$ for 
two successive shells $\ell$ and $\ell+1$ [Eq.(\ref{Eq.fl1})].   
Using this equation we construct a correlation function
$c(r)$ [Eq.(\ref{Eq.rg})] of the network which characterizes the 
quality of the network connectedness. 
We apply this measure for several model and real networks. We find that human 
collaboration networks are
usually poorly-connected compared to the random networks with the 
same degree distribution. The same is true for the WS small-world model. 
In contrast, we find that the Internet is a well-connected network.
The same is true for the BA model. Thus our results indicate that the WS
model and the BA model correctly reproduce 
an essential feature of the real-world models
they were designed to mimic, namely, social networks and the Internet, respectively. 
Finally we apply
Eq. (\ref{Eq.fl1}) to derive the power law distribution of 
the number of nodes in the shells with $\ell>>d$ \cite{Shao}. 

\section{Acknowledgments} 
We wish to thank the ONR, EU project
Epiwork, the Israel Science Foundation for financial support.
S.V.B. thanks the Office of the Academic Affairs of Yeshiva
University for funding the Yeshiva University high performance
computer cluster and acknowledges the partial
support of this research through the Dr. Bernard W. Gamson
Computational Science Center at Yeshiva College.

\appendix
\section{Differential equations for $\Lambda(r)$ and $\Lambda_{\ell}(r)$} \label{A1}

\begin{figure*}[h!]
    \includegraphics[width=11cm,angle=0]{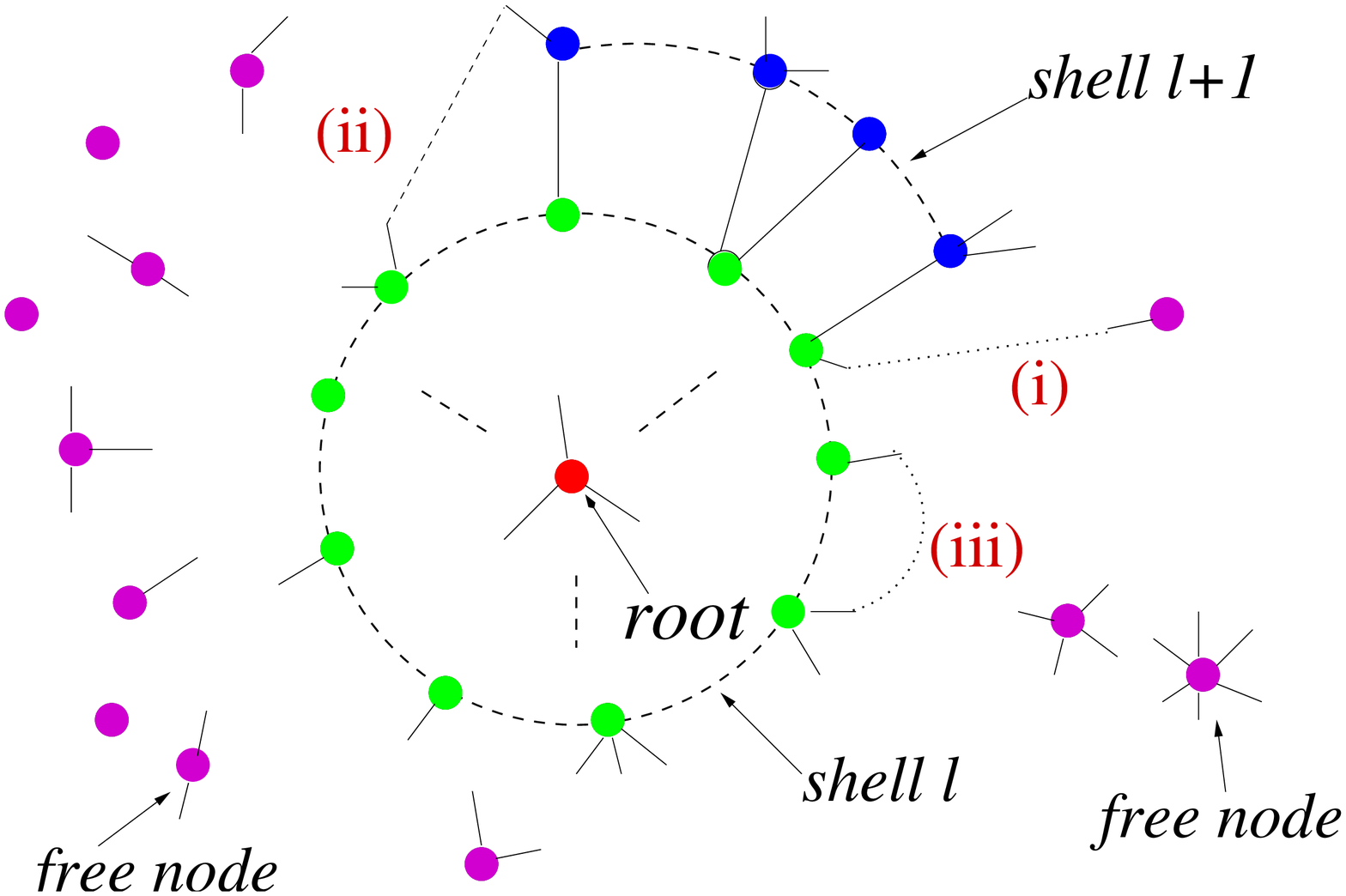}
    \caption{ (color online) Building the network begins from a randomly
chosen node (root), shown in red at the center of the figure.
This schematic illustration shows the network during the building of shell $\ell+1$.
We do not start to build shell $\ell+1$ until shell $\ell$ is
completed. All the nodes which are already included in shell $\ell+1$ are
shown in blue, while the free nodes not yet connected in shell $\ell+1$ are shown in purple.
At a certain time step, 
in order to connect an open link from shell $\ell$ to another
open link, we must consider three scenarios: (i) Connecting to an open 
link taken from
a free node. (ii) Connecting to an open link 
from shell $\ell+1$. (iii) Connecting
to another open link from shell $\ell$. This way the aggregate keeps
growing shell after shell until all the open links are connected. Note that in
scenarios (ii) and (iii) there
is a chance to create parallel links (two links connecting a
pair of nodes) and circular links (one link with two ends connected to the same node).
For a large network with a finite $\tilde k$, such events occur with a negligible probability.
}
\label{fig1}
\end{figure*}

\begin{figure}[h!]
      \includegraphics[width=6cm,angle=-90]{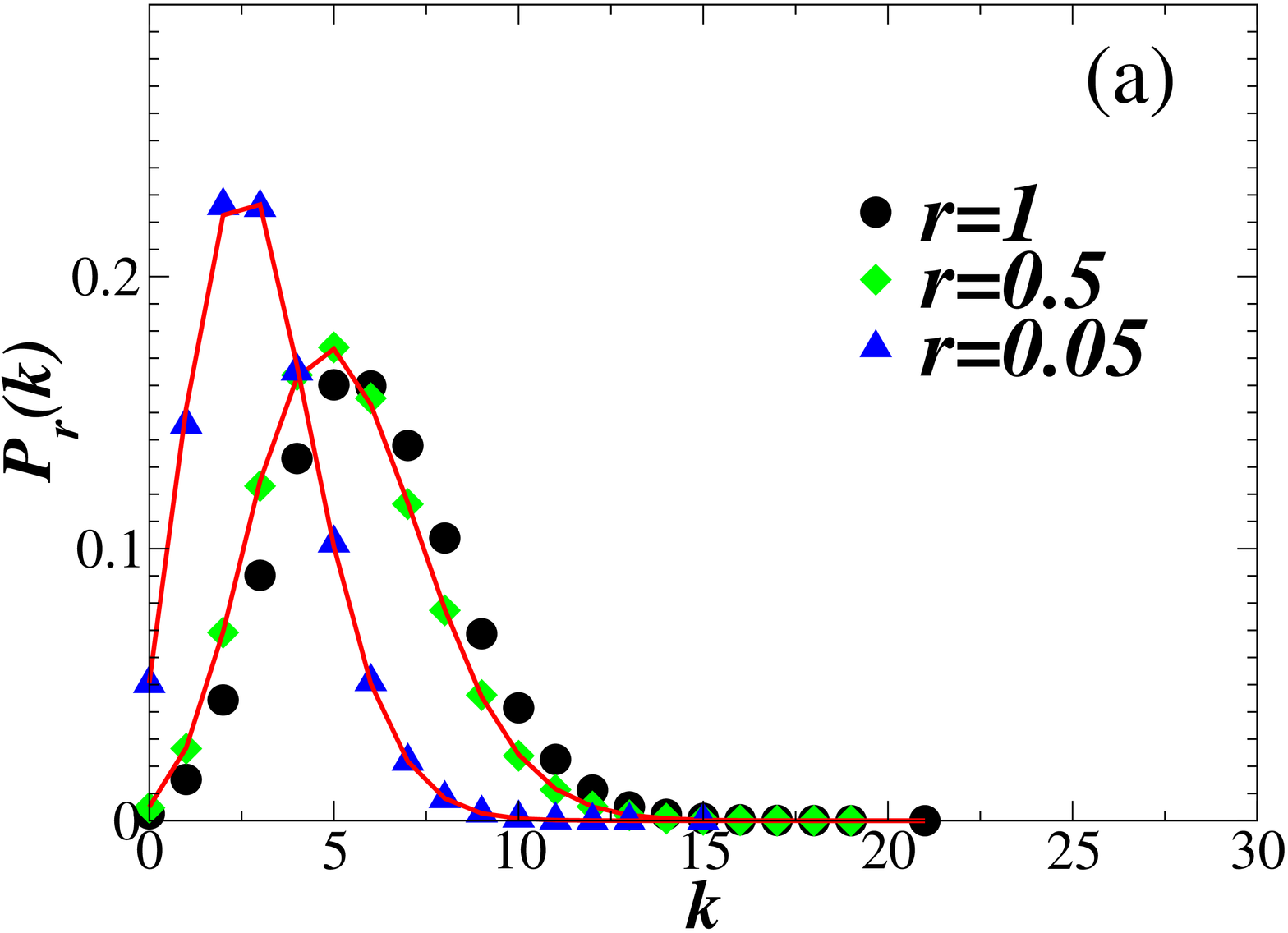}
        \includegraphics[width=6cm,angle=-90]{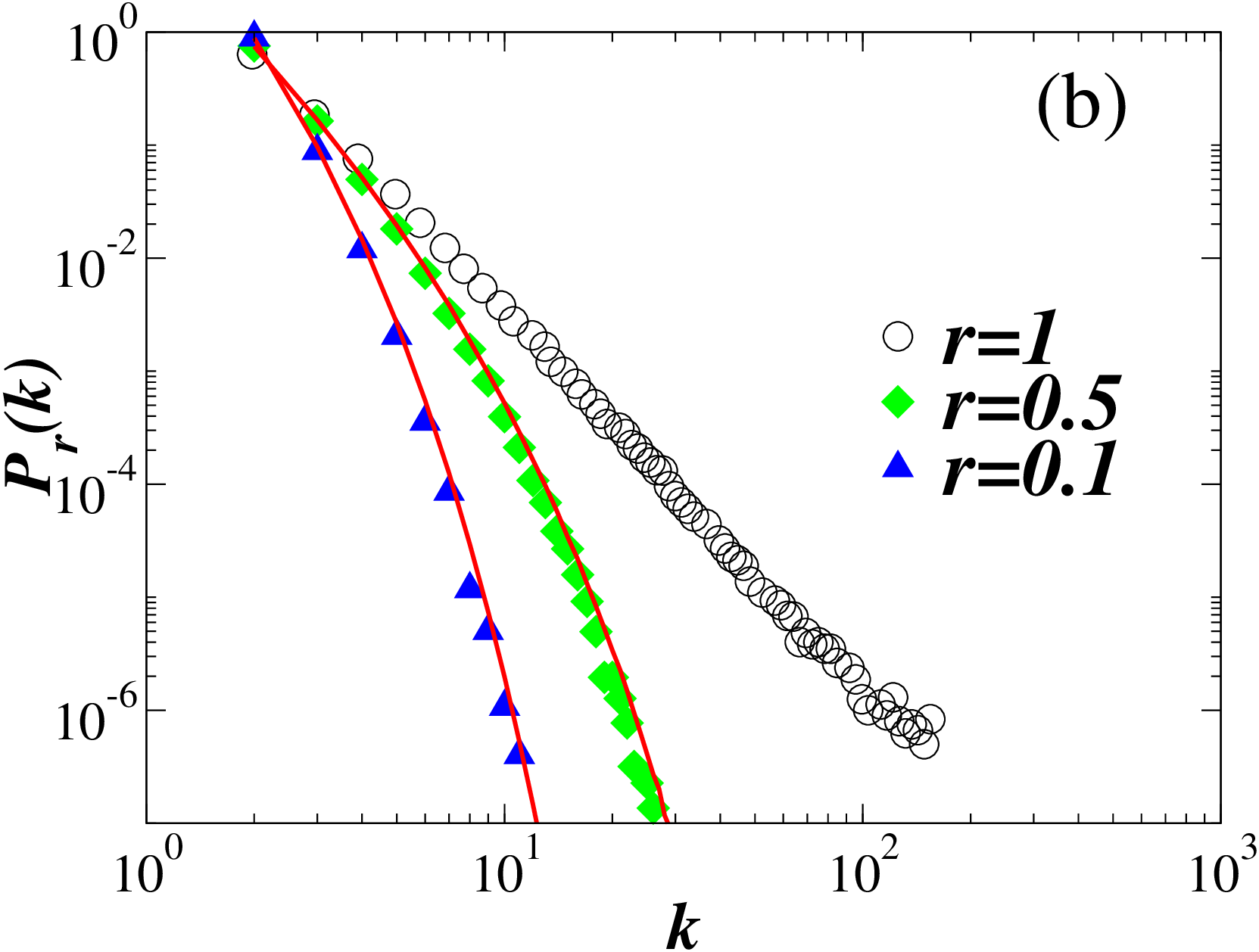}
      \caption{ The degree distribution, $P_r(k)$, in $E_r$ 
for (a) an ER network with $N=10^6$, $\langle k \rangle=6$
and $r=1, 0.5$ and $0.05$.
The simulation results (symbols)
agree very well with the 
theoretical predictions (lines) of Eq.(\ref{Eq.Pkf}).
(b) a SF network with $\lambda=3.5$, $k_{\rm min}=2$ and $N=10^6$, 
$P_r(k)$ with
$r=1, 0.5$ and $0.1$. The simulation results shown by symbols
fit well with 
the theoretical predictions of Eq.(\ref{Eq.Pku}). For a SF network, we
compute Eq.(\ref{Eq.Pku}) numerically using the $P(k)$ obtained from the 
generated network.
}
\label{fig2}
\end{figure}

\begin{figure}[h!]
       \includegraphics[width=6cm,angle=-90]{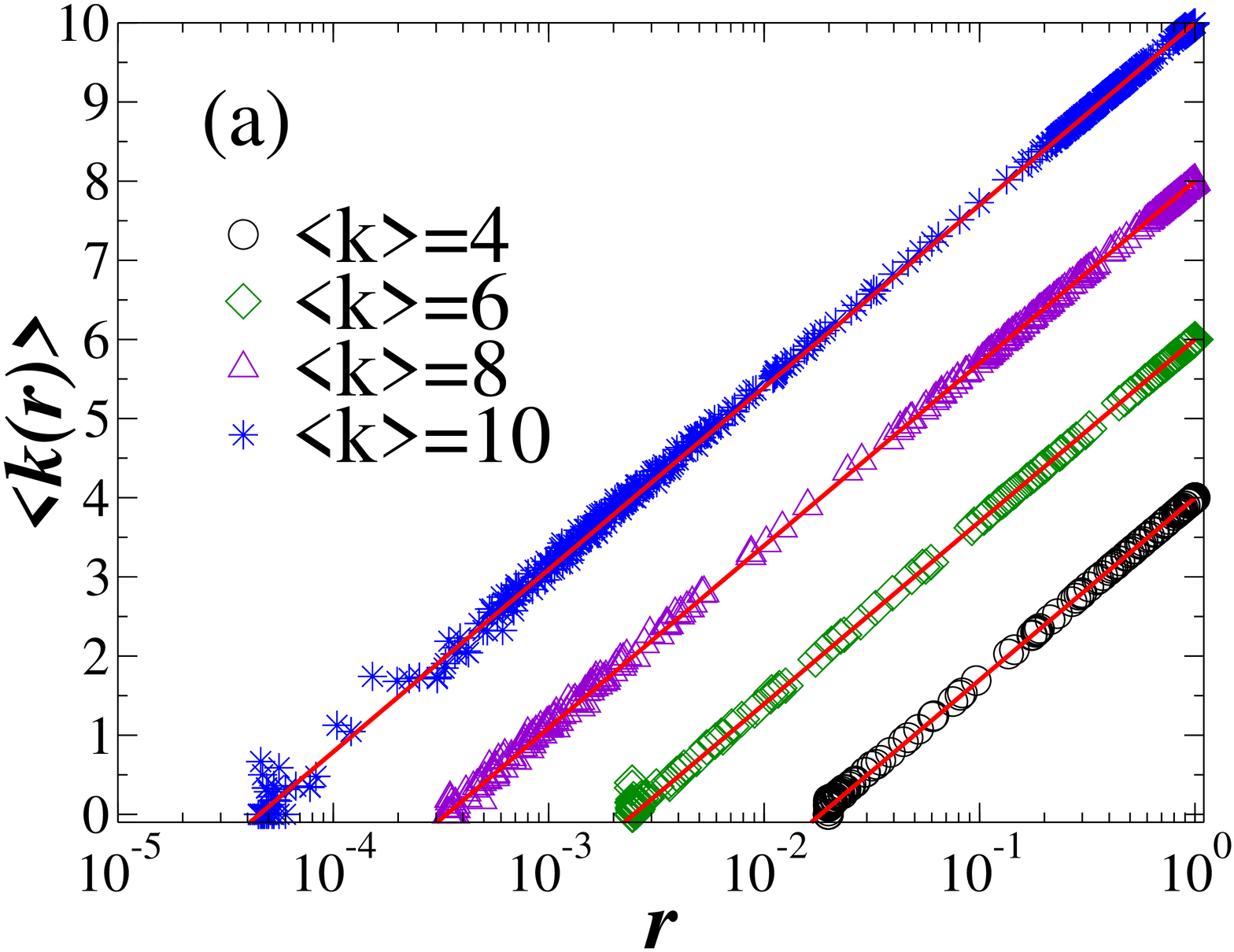}
       \includegraphics[width=6cm,angle=-90]{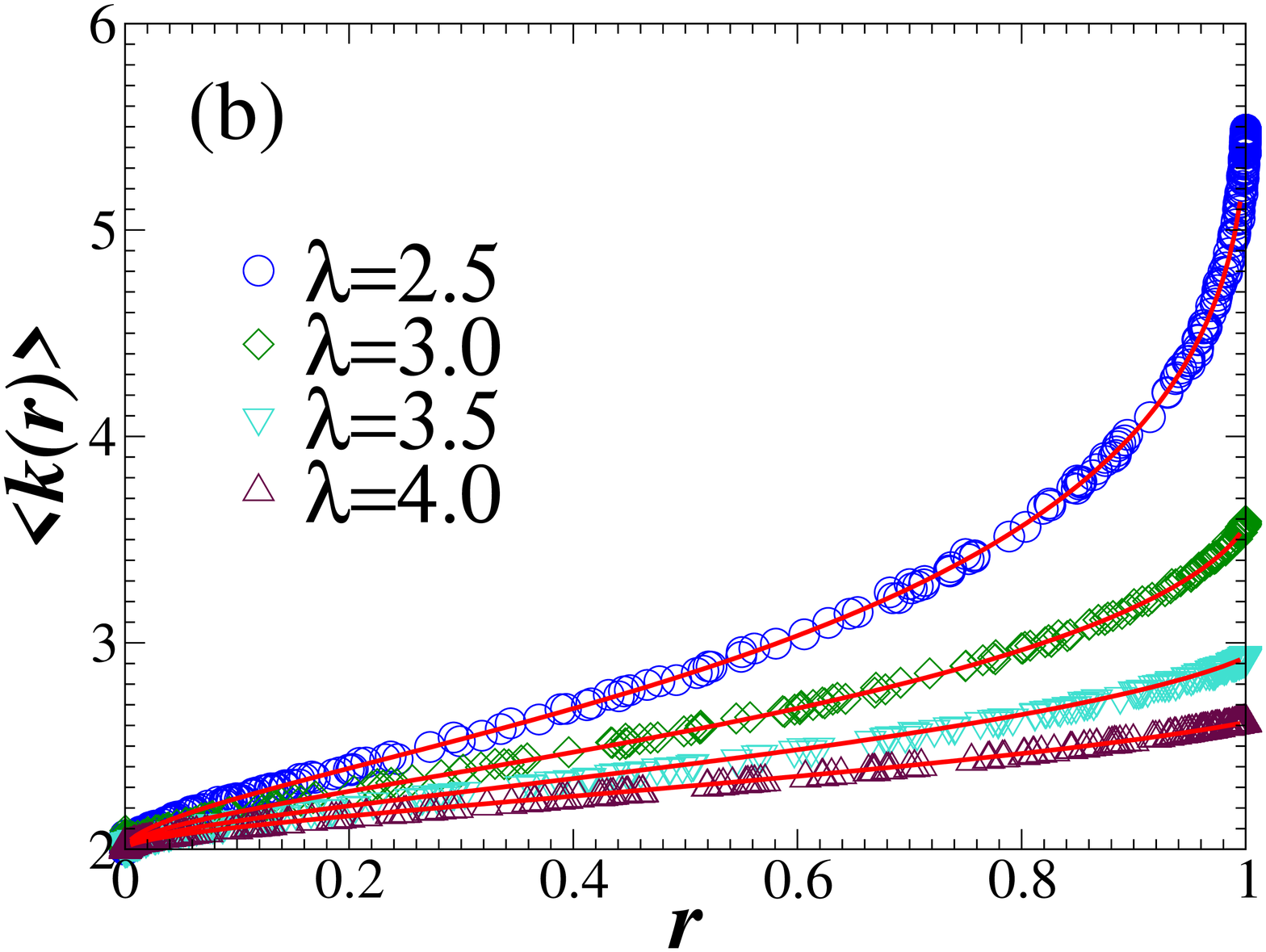}
      \caption{Average degree $\langle k(r) \rangle$  of the nodes in $E_r$ 
 as a function of $r$ for (a) four ER networks with different values of
$\langle k \rangle$, and (b) four SF networks with $k_{\rm min}=2$ and 
different values of $\lambda$. 
The symbols represent the simulation results
for ER and SF networks of size $N=10^6$.
The lines in (a) represent Eq.(\ref{Eq.kf}). The lines in (b) are the
numerical results of Eq.(\ref{Eq.ku}), using the degree distribution 
obtained from the networks. 
}
\label{fig3}
\end{figure}

\begin{figure*}[h!]
     \includegraphics[width=6cm,angle=-90]{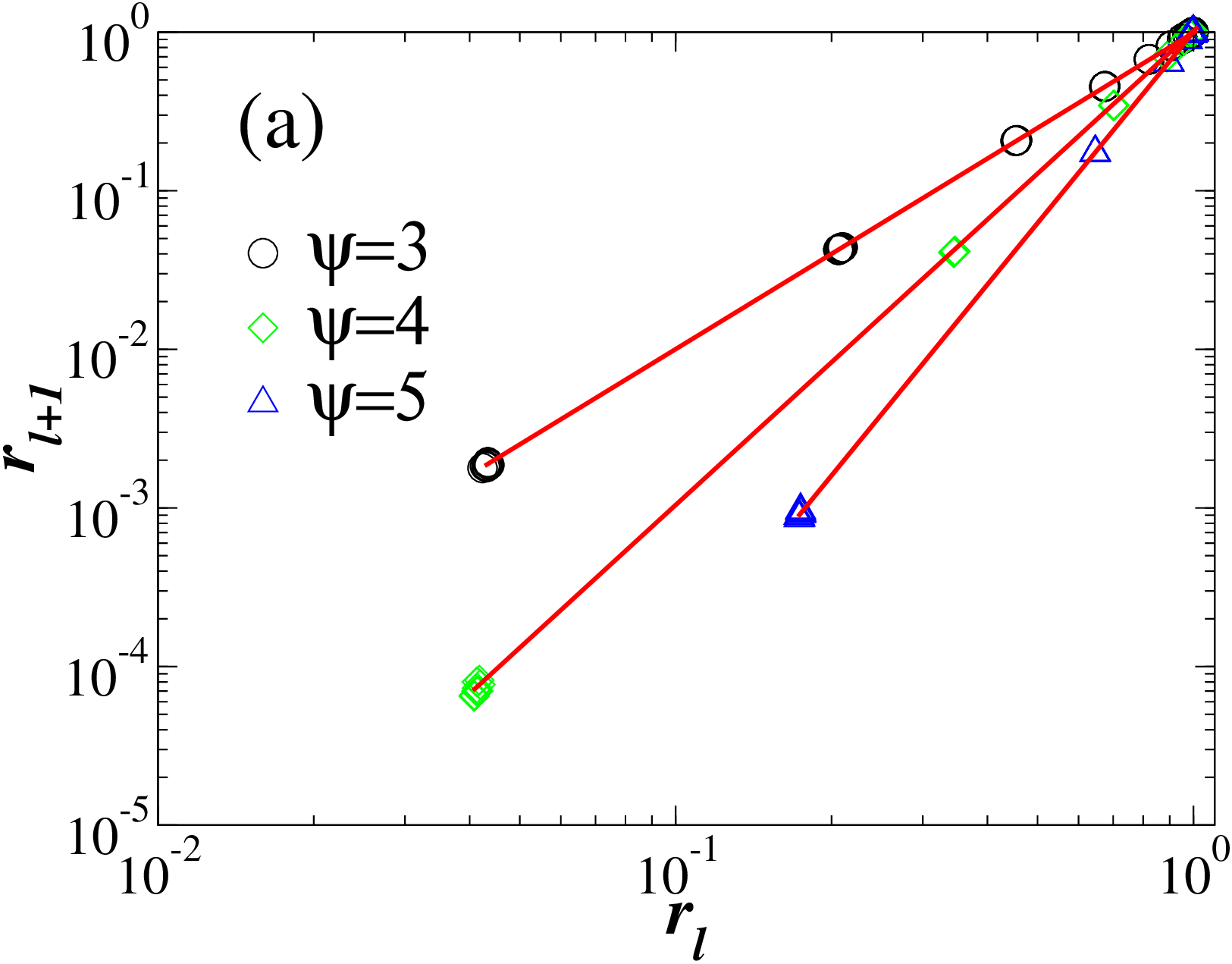}
    \includegraphics[width=6cm,angle=-90]{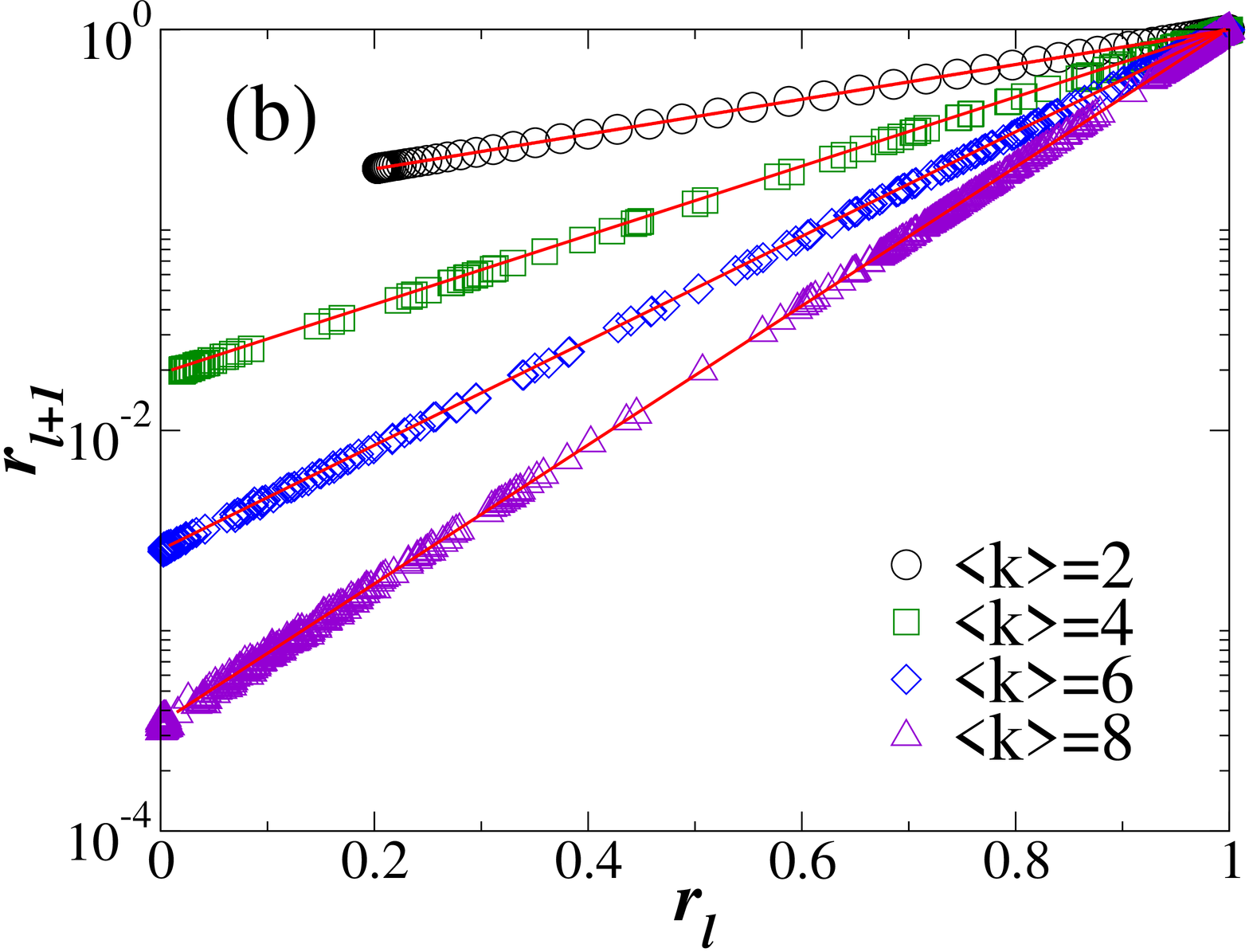}
    \includegraphics[width=6cm,angle=-90]{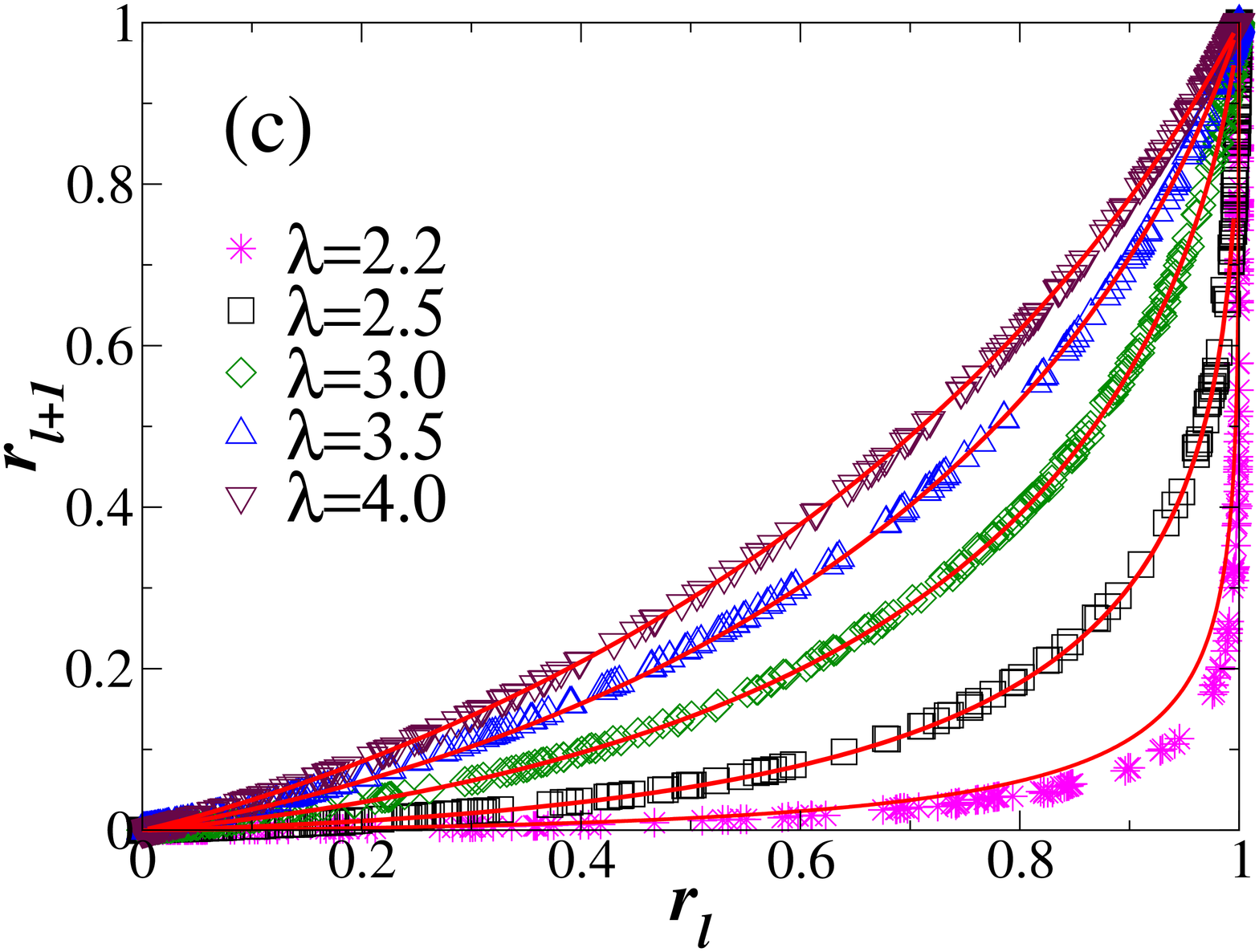}
     \includegraphics[width=6cm,angle=-90]{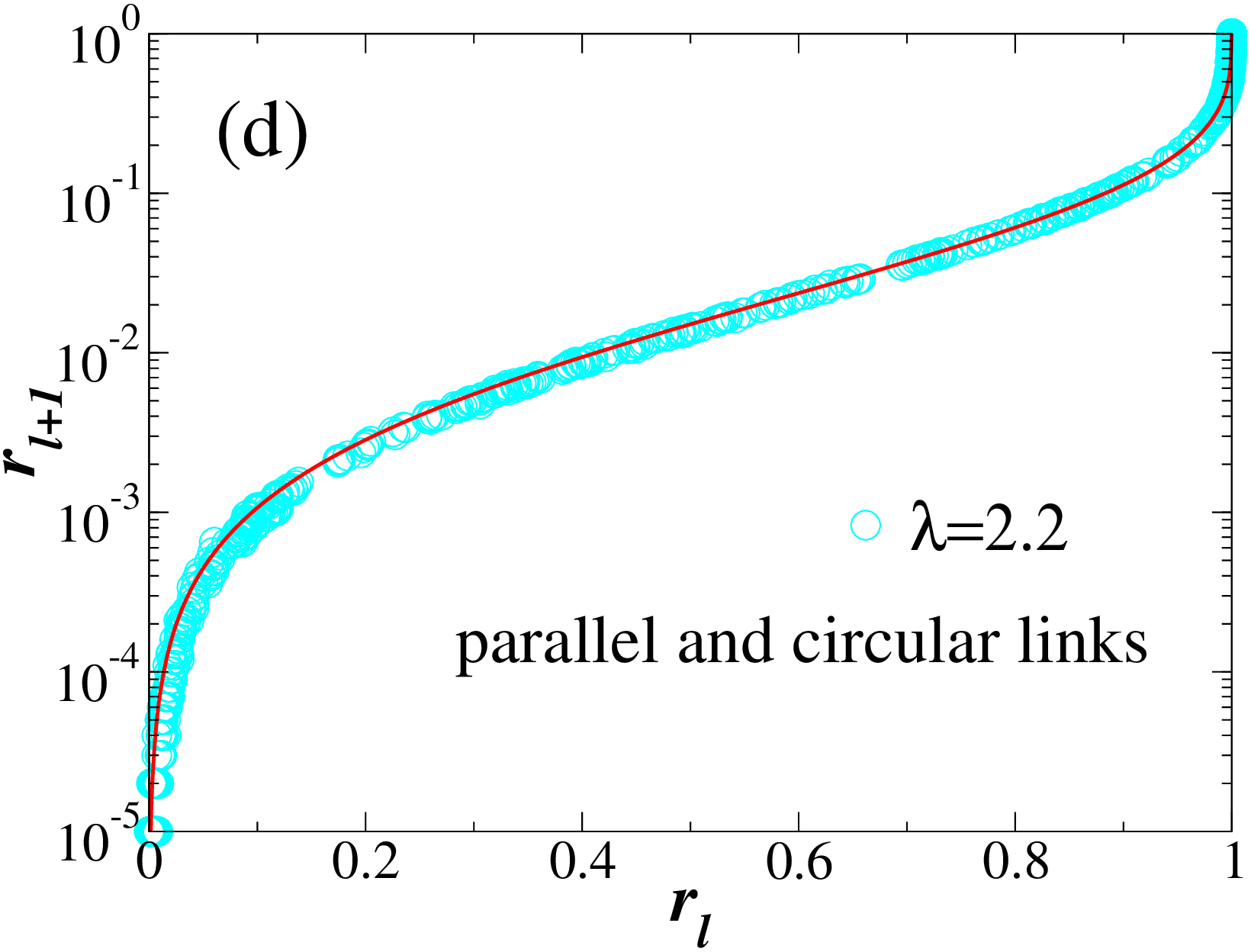}
    \caption{ (Color online) For a randomly chosen root in the network, 
the fraction of nodes $r_{\ell+1}$
in $E_{\ell+1}$ as a function of the the fraction of 
nodes $r_{\ell}$ in $E_{\ell}$ for
(a) three RR networks of size $N=10^5$ with different $\psi$.
The red lines represents the theoretical prediction of Eq.(\ref{Eq.ffrl}).
(b) Four ER networks of size $N=10^5$ with different $\langle k \rangle$.
The red lines represent the theoretical predictions of Eq.(\ref{Eq.ffa}).
(c) Five SF networks of size $N=10^5$ with different values of $\lambda$. 
The red lines shown are the
numerical results of Eq.(\ref{Eq.fl1}) using the degree distribution obtained 
from the simulation. 
For $\lambda\geq 2.5$, the agreement between the theory [Eq.(\ref{Eq.fl1})]
and the simulation results is perfect.
(d) A SF network of size $N=10^5$ with $\lambda=2.2$, which allows 
parallel and circular 
links (PCL) during its construction.  
Simulation results of SF networks 
with PCL show excellent agreement with the theory (full line). 
}
\label{fig4}
\end{figure*}

\begin{figure*}[h!]
       \includegraphics[width=6cm,angle=-90]{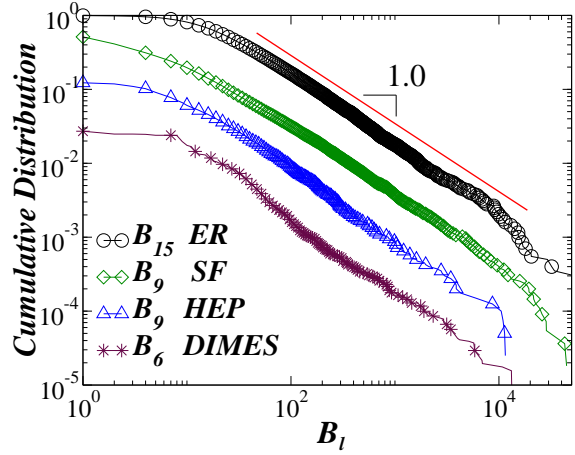}
       \caption{ Cumulative distribution function of $P(B_{\ell})$, of 
the number of nodes $B_\ell$
in shell $\ell$ ($\ell\gg d$) for an ER network with $\langle k \rangle=4$, $N=10^6$ and $d\approx 10.0$, 
a SF network with $ \lambda=2.5$, $N=10^6$ and $d\approx 4.7$, the HEP network 
($d\approx 4.2$) and the DIMES network ($d\approx 3.3$). 
Note that slope $-1$ of the cumulative distribution function implies
$P(B_{\ell})\sim B^{-2}_{\ell}$, which holds for all four examples, as well as 
for many other networks studied \cite{Shao}.
}
\label{figab}
\end{figure*}

\begin{figure*}[h!]
       \includegraphics[width=6cm,angle=-90]{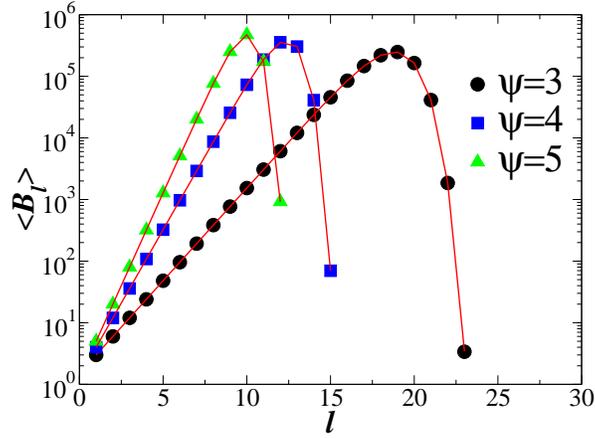}
      \caption{ The average number of nodes,
$\langle B_\ell \rangle$, in shell $\ell$ as a function of the 
shell index $\ell$
for the RR network with different $\psi$. The theoretical predictions (full lines) calculated from
Eq.(\ref{Eq.bll}) and Eq.(\ref{Eq.ffrlm}) fit very well 
the simulation results (symbols).
}
\label{fig7}
\end{figure*}

\begin{figure}[h!]
    \includegraphics[width=6cm,angle=-90]{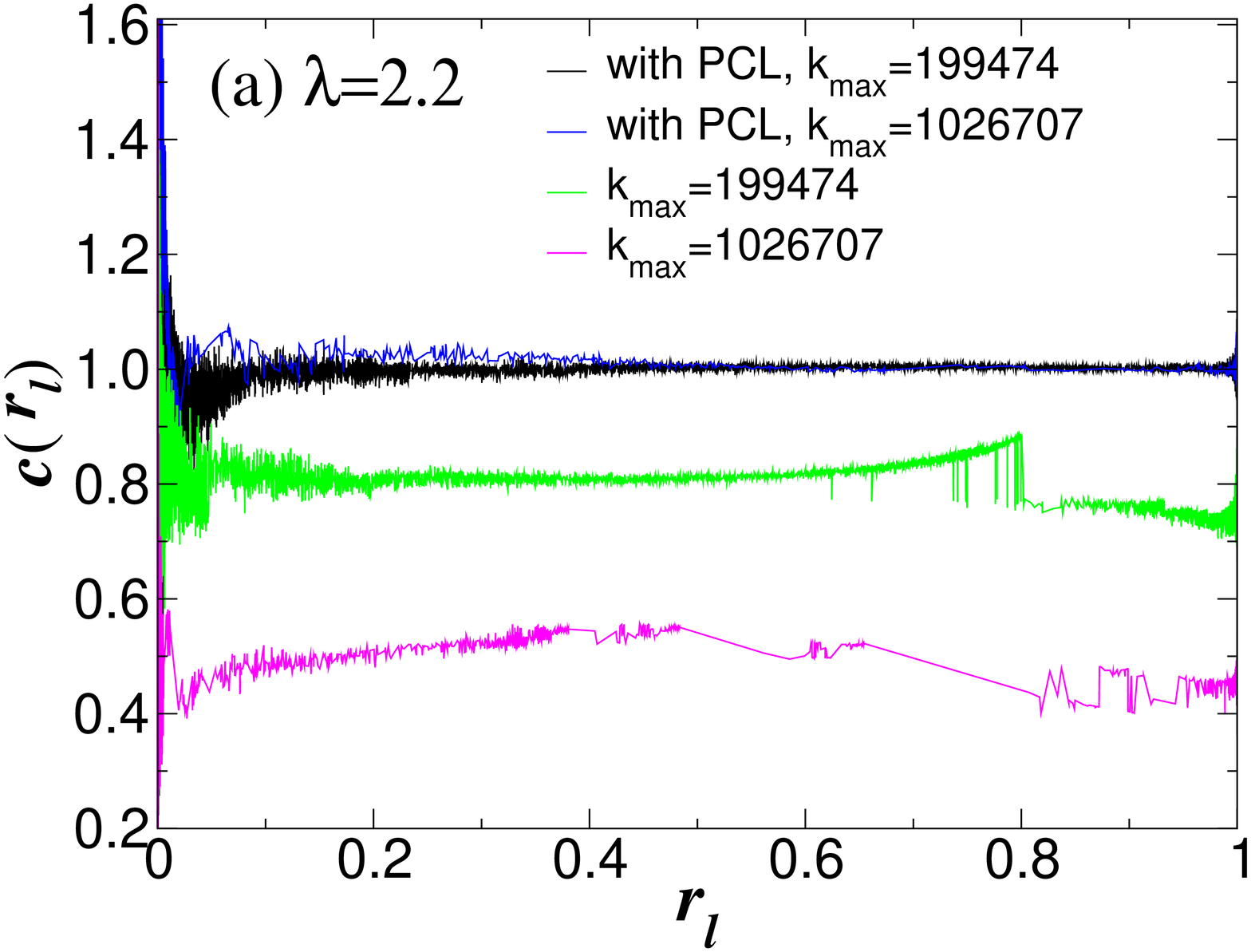}
     \includegraphics[width=6cm,angle=-90]{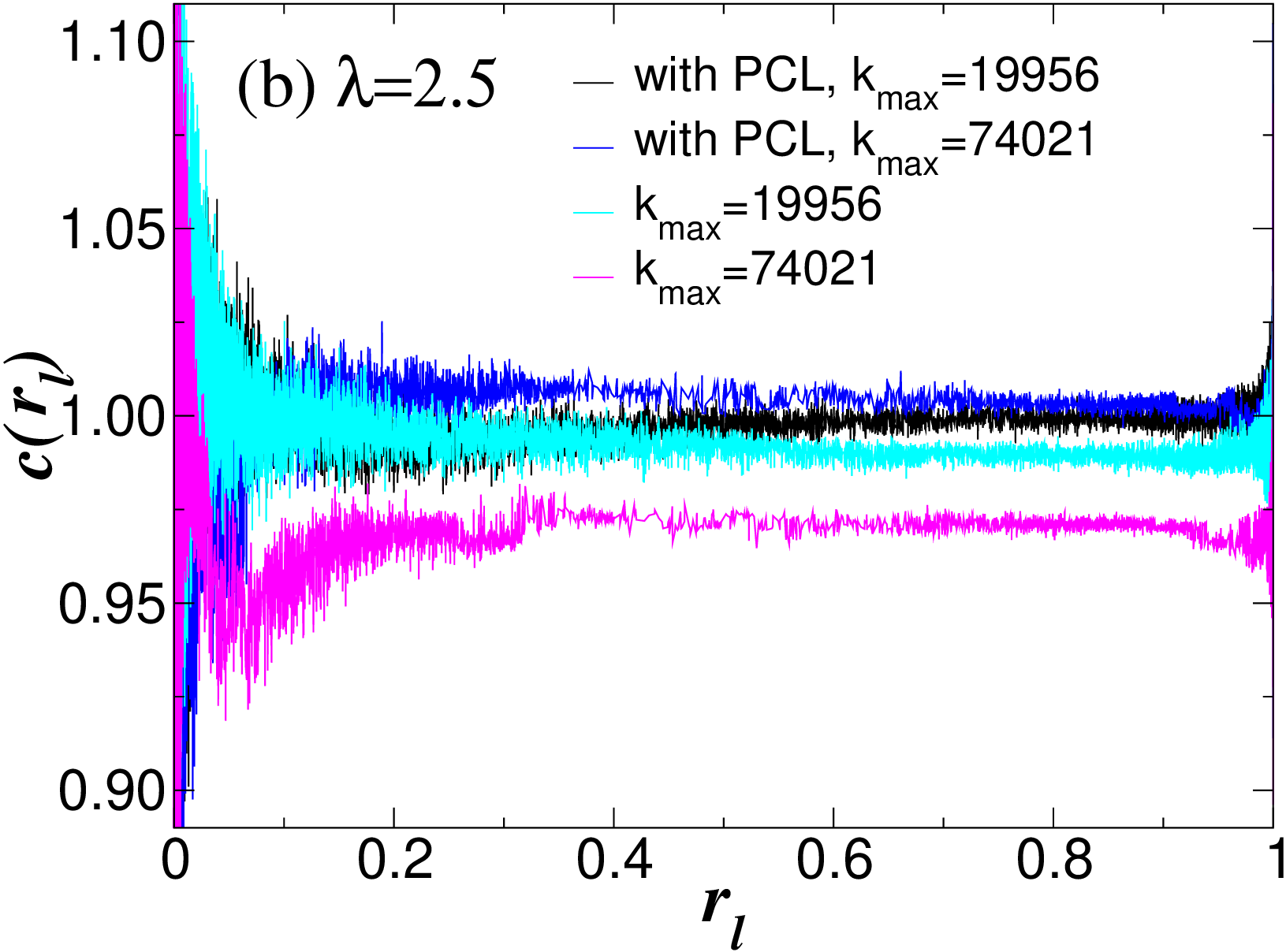}
           \caption{$c(r_{\ell})$ for SF networks. (a) The case  
$\lambda=2.2$. Values of $k_{\rm max}$ equal and larger than the 
natural cutoff ($k_{\rm max}=k_{\rm min}N^{1/(\lambda-1)}\approx 2\times 10^5$)
are compared for networks with and without parallel and circular links  
(PCL). Notice that, the 
discontinuity of the lines is due to the existence of the large degree nodes.
(b) The case $\lambda=2.5$. Similarly, values of $k_{\rm max}$ equal and 
larger than the natural cutoff ($k_{\rm max}\approx 2\times 10^4$)
are compared for networks with and without PCL.  
}
\label{fig6a}
\end{figure}

\begin{figure}[h!]
     \includegraphics[width=6cm,angle=-90]{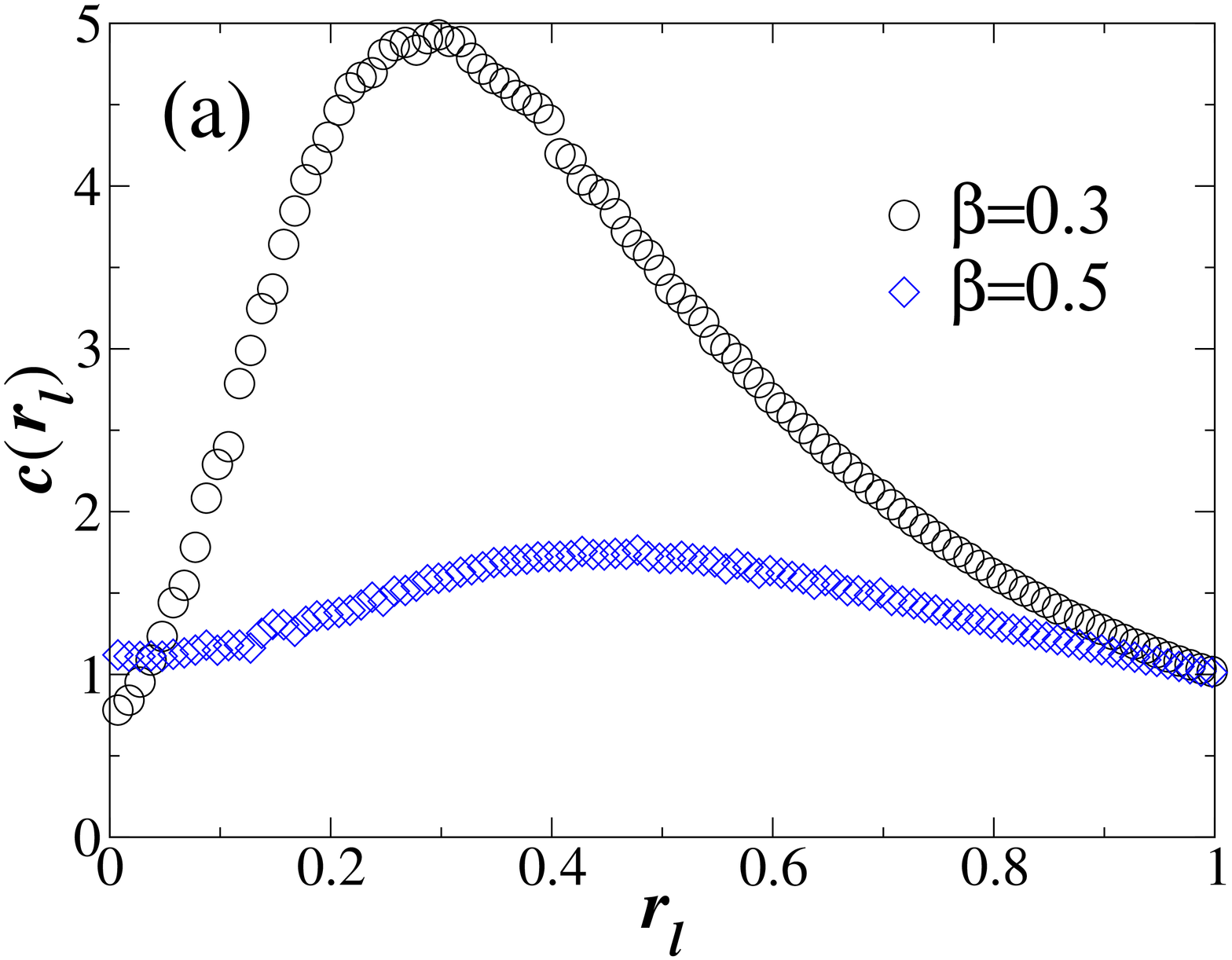}
     \includegraphics[width=6cm,angle=-90]{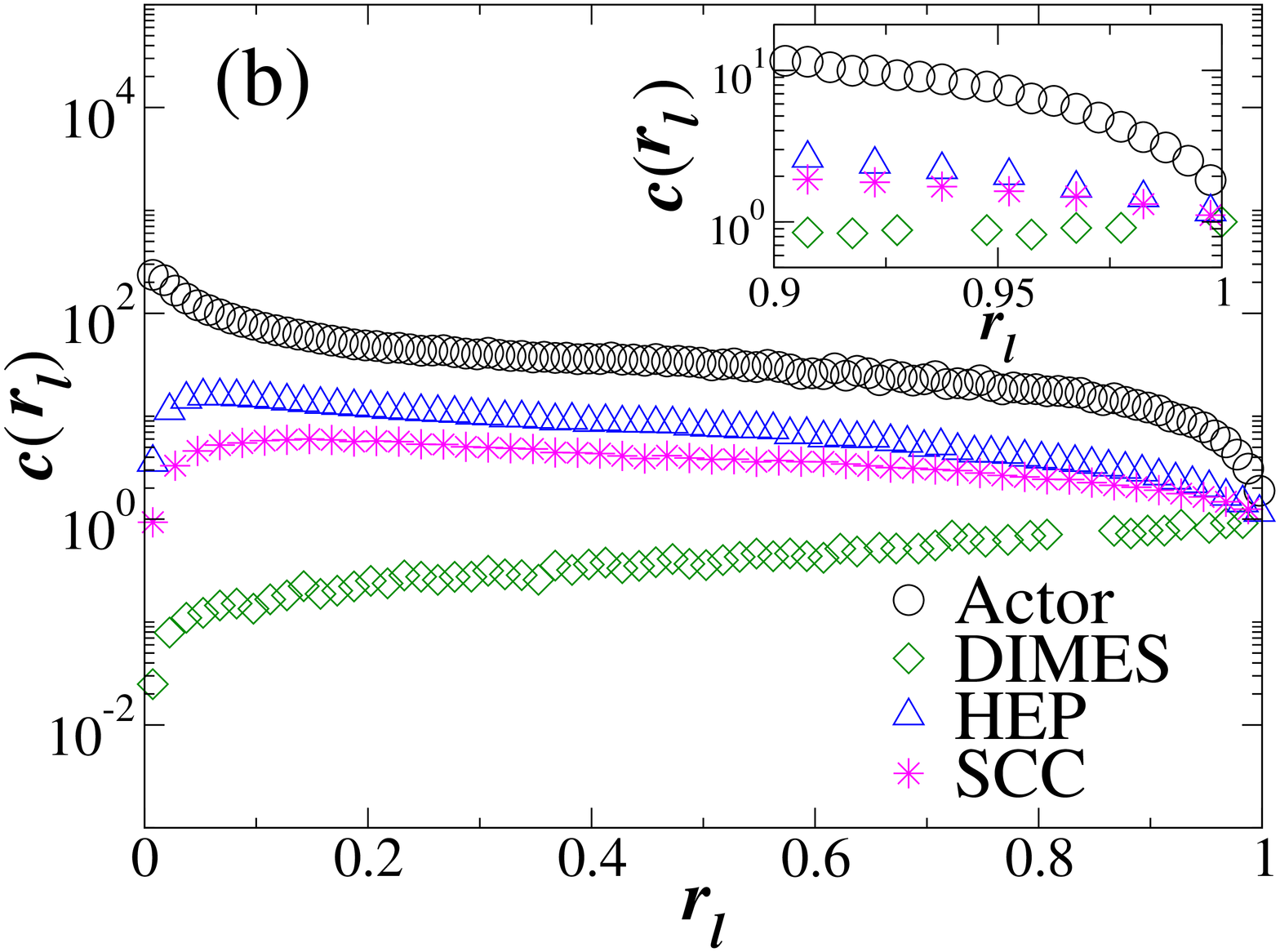}
     \includegraphics[width=6cm,angle=-90]{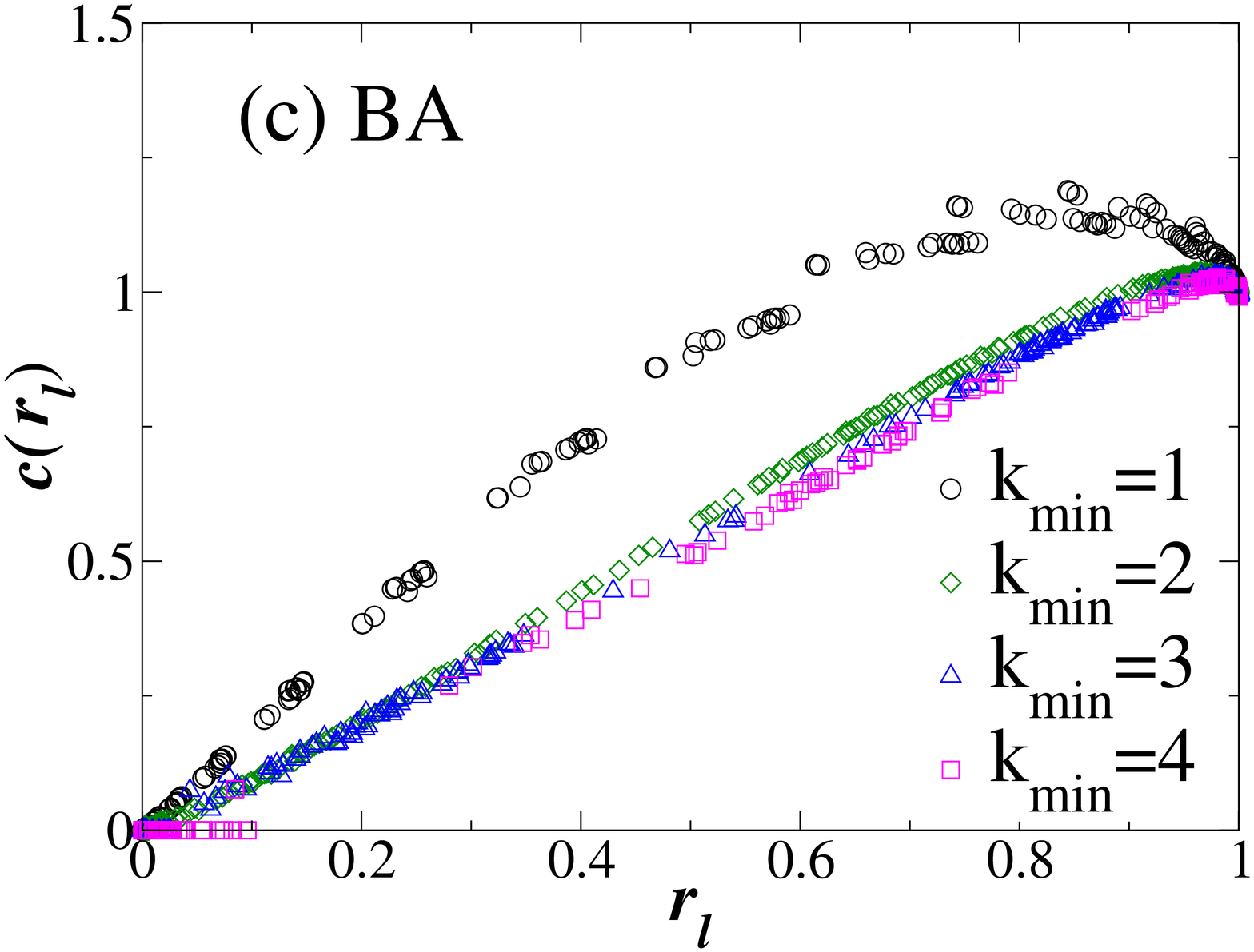}
           \caption{$c(r_{\ell})$ for various networks. 
(a) WS network with $\psi=4$ and $\beta=0.3$ and 0.5. 
(b) Four real networks: Actor collaboration
network (Actor), High Energy Physics citations network (HEP), AS Internet network (DIMES),
and Supreme Court Citation network (SCC). In the insert we show the enlarged 
area of $r>0.9$. (c) BA networks of size $N=10^6$ with different $k_{\rm min}$.
}
\label{fig6}
\end{figure}

In this appendix, 
we derive the differential equations (Eq. (\ref{Eq.dequations1})) 
for $\Lambda(t)$ and $\Lambda_{\ell}(t)$.
At time $t$, the total number of open links in the r-exterior $E_r$ of 
the unconnected
nodes is
$rN \langle k(r) \rangle$. 
At step $t$, we connect one open link from the aggregate to another open link. There
is a probability 
$$
\frac{r(t) \langle k(r(t)) \rangle}{r(t)\langle k(r(t)) \rangle+\Lambda(t)}
$$ 
that will be connected to a free node.
Thus,
\begin{equation}\label{Eq.nf}
Nr(t+1)=Nr(t)-\frac{r(t) \langle k(r(t)) \rangle}{r(t)\langle k(r(t)) \rangle+\Lambda(t)}.
\end{equation}
To derive the differential equations for $\Lambda(t)$, we need to consider 
all three different scenarios which we illustrated in Fig.~\ref{fig1}.
If we connect an open link from the aggregate to a node which is not yet connected to the aggregate (scenario (i) in Fig.~\ref{fig1}),
on average $\Lambda(t)$ will increase by $\tilde{k}(r(t))/N$. If we connect the open link from the aggregate to another open link either
from shell $\ell$ or shell $\ell+1$ (scenarios (iii) and (ii) in Fig.~\ref{fig1}), $\Lambda(t)$ will decrease by $1/N$.
Because we connect links at random, the probability
of scenario (i) is 
$$
\frac{r(t)\langle k(r(t)) \rangle}{r(t)\langle k(r(t)) \rangle+\Lambda(t)}
$$
and the probability of scenarios (ii) or (iii)
is 
$$
\frac{\Lambda(t)}{r(t)\langle k(r(t)) \rangle+\Lambda(t)}.
$$
Thus, we can write down the evolution of $\Lambda(t)$ as
\begin{equation}\label{Eq.ltp}
\Lambda(t+1)=\Lambda(t)-\frac{1}{N}+\frac{\tilde{k}(r(t))}{N} \frac{r(t)\langle k(r(t)) \rangle}{r(t)\langle k(r(t)) \rangle+\Lambda(t)}
-\frac{1}{N} \frac{\Lambda(t)}{r(t)\langle k(r(t)) \rangle+\Lambda(t)}.
\end{equation}
For $N \to \infty$, Eqs. (\ref{Eq.nf}) and (\ref{Eq.ltp}) lead respectively to
\begin{equation}\label{Eq.ndf}
\frac{dr(t)}{dt}=-\frac{1}{N}\frac{r(t)\langle k(r(t)) \rangle}{r(t)\langle k(r(t))\rangle+\Lambda(t)},
\end{equation}
and
\begin{equation}\label{Eq.dltp}
\frac{d\Lambda(t)}{dt}=-\frac{1}{N}+\frac{\tilde{k}(r(t))}{N}\frac{r(t)\langle k(r(t))
    \rangle}{r(t)\langle k(r(t)) \rangle +\Lambda(t)}-
     \frac{1}{N}\frac{\Lambda(t)}{r(t)\langle k(r(t)) \rangle+\Lambda(t)}.
\end{equation}
Dividing Eq.(\ref{Eq.dltp}) by Eq.(\ref{Eq.ndf}) we obtain the differential equation for $\Lambda$ as a function of $r$
\begin{equation}\label{Eq.dlfp}
\frac{d\Lambda(r)}{dr}=-\tilde{k}(r)+1+\frac{2\Lambda(r)}{r\langle k(r)\rangle}.
\end{equation}

$\Lambda_{\ell}(t)$ behaves similarly to $\Lambda(t)$ except that we only need to
consider the effect of scenario (iii) of Fig.~\ref{fig1}. Accordingly, 
the evolution of $\Lambda_{\ell}$ can
be written as
\begin{equation}\label{Eq.lltp}
\Lambda_{\ell}(t+1)=\Lambda_{\ell}(t)-\frac{1}{N}
-\frac{1}{N} \frac{\Lambda_{\ell}(t)}{r(t)\langle k(r(t)) \rangle+\Lambda(t)},
\end{equation}
which for $N\to \infty$ is
\begin{equation}\label{Eq.dlltp}
\frac{d\Lambda_{\ell}(t)}{dt}=-\frac{1}{N}-\frac{1}{N}\frac{\Lambda_{\ell}(t)}{r(t)\langle k(r(t)) \rangle+\Lambda(t)}.
\end{equation}
Dividing Eq.(\ref{Eq.dlltp}) by Eq.(\ref{Eq.ndf}), we get
\begin{equation}\label{Eq.dllfp}
\frac{d\Lambda_{\ell}(r)}{dr}=1+\frac{\Lambda(r)}{r\langle k(r) \rangle}+\frac{\Lambda_{\ell}(r)}{r\langle k(r) \rangle}.
\end{equation}

\section{Solution of $G_{1}(f_{\infty})=f_{\infty}$ and $\delta=-\ln G_{1}'(f_\infty)/\ln\tilde{k}$}\label{B1}

\begin{figure}[h!]
       \includegraphics[width=6cm,angle=-90]{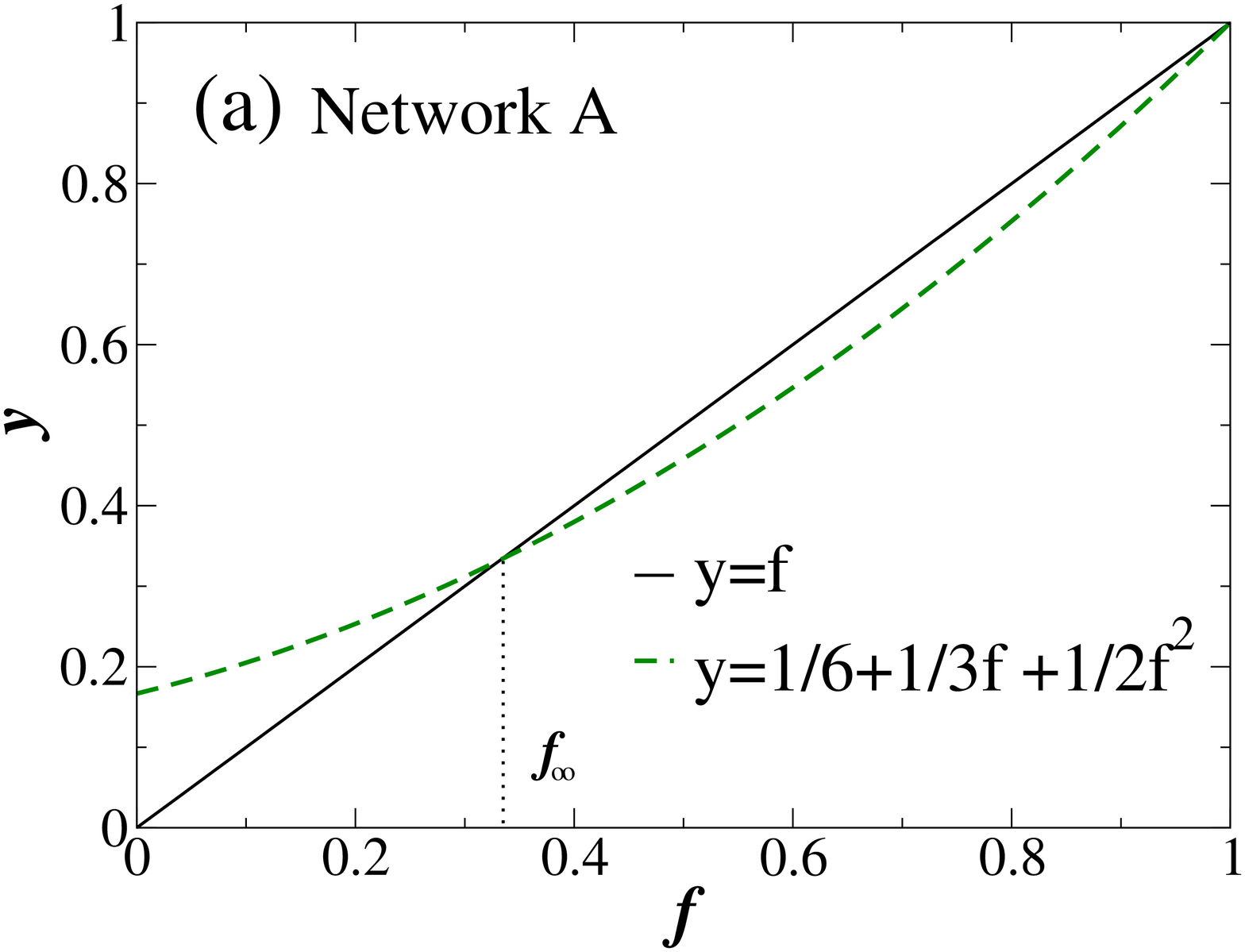}
       \includegraphics[width=6cm,angle=-90]{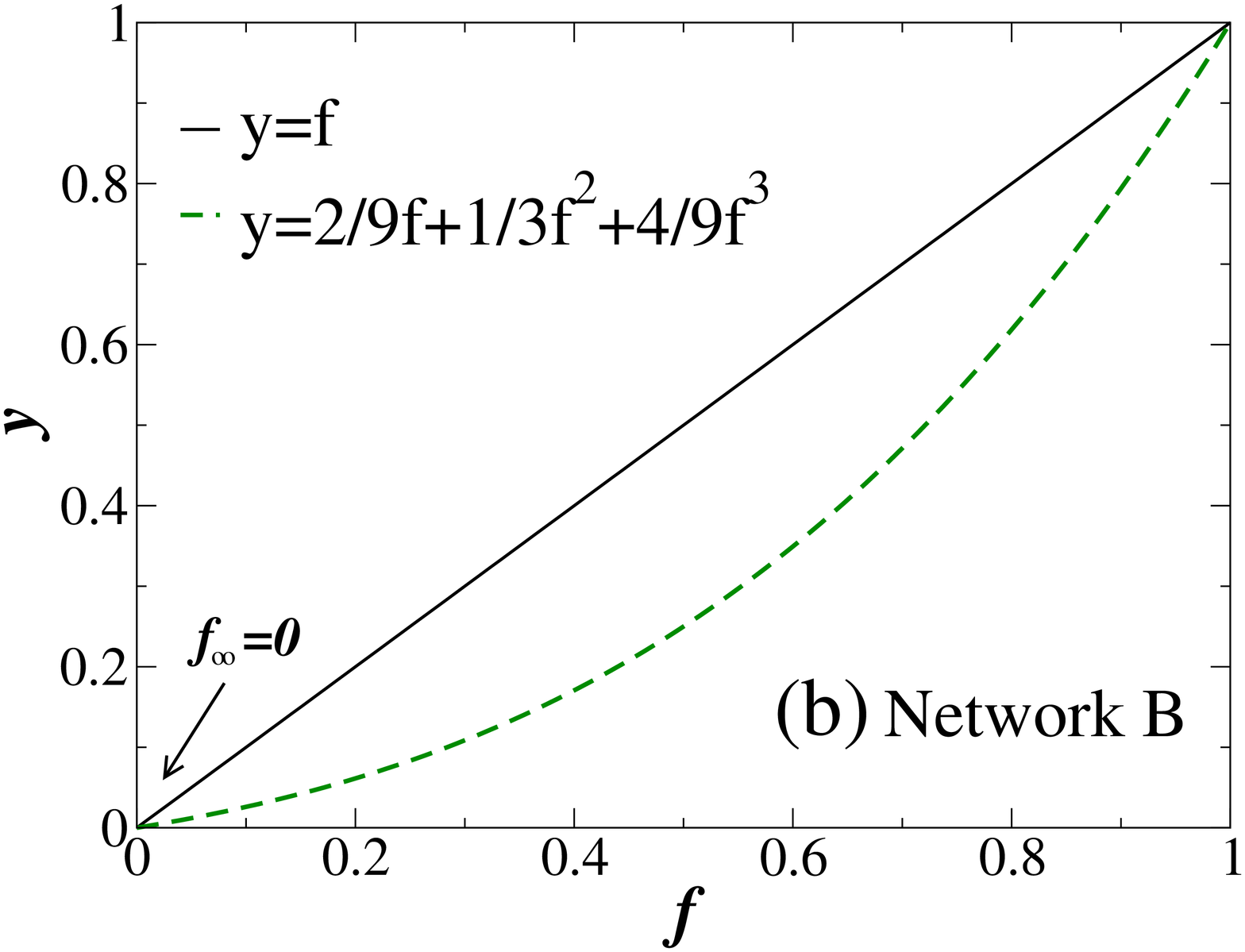}
      \caption{ Plots of both sides of Eq.(\ref{Eq.uinfty}) for 
(a) Network A, with 
equal probability of having
degree 1, 2 and 3, and (b) Network B, with equal probability of having
degree 2, 3 and 4.
For network A, a non-zero solution $f_\infty$ can be seen. For network B, 
$f_\infty=0$ is the solution of Eq.(\ref{Eq.uinfty}).
}
\label{figf}
\end{figure}

The numerical solutions of $G_{1}(f_{\infty})=f_{\infty}$ can be shown
by a simple example. Suppose we have three simple
networks A, B and C. In network A,
all the nodes can only have degree 1, 2 and 3.
In network B, the degree can be 2, 3 and 4.  
In network C,
the degree can be 3, 4 and 5. For all three examples, the probability of each degree
is 1/3.
We can write $G_0$ and $G_1$ for three network as
\begin{eqnarray}
   G_{0,A}(x)&=\frac{1}{3}x+\frac{1}{3}x^2+\frac{1}{3}x^3 \label{Eq.g0A}\\
   G_{0,B}(x)&=\frac{1}{3}x^2+\frac{1}{3}x^3+\frac{1}{3}x^4 \label{Eq.g0B}\\
   G_{0,C}(x)&=\frac{1}{3}x^3+\frac{1}{3}x^4+\frac{1}{3}x^5 \label{Eq.g0C}
\end{eqnarray}
The average degrees $\langle k\rangle=G_0^{'}$ of A, B and C are 2, 3 and 4 respectively. Using 
the above expressions for $G_0$ we can construct the expressions for $G_1(x)$:
\begin{eqnarray}
   G_{1,A}(x)&=\frac{1}{6}+\frac{1}{3}x+\frac{1}{2}x^2 \label{Eq.g1A}\\
   G_{1,B}(x)&=\frac{2}{9}x+\frac{1}{3}x^2+\frac{4}{9}x^3 \label{Eq.g1B}\\
   G_{1,C}(x)&=\frac{1}{4}x^2+\frac{1}{3}x^3+\frac{5}{12}x^4 \label{Eq.g1C}
\end{eqnarray}
The branching factors $\tilde{k}=G_1^{'}(1)$ of A, B and C are 2/3, 20/9 and 19/6 respectively.
The numerical solutions of $G_{1}(f_{\infty})=f_{\infty}$ for network A
and B is shown in Figs.~\ref{figf}a and \ref{figf}b, where we plot 
the functions $y=f$ and $y=G_1(f)$ on the same plot.
From Fig.~\ref{figf}, we can see that there is a
non-zero $f_\infty=1/3$ for network A and $f_\infty=0$ for network B.
For network C, we also have $f_\infty=0$.
Whether
we can have a non-zero $f_\infty$ depends on
the first term of $G_1(x)$, which depends on $P(k=1)$, the
probability of having nodes with degree 1. If $P(k=1)\neq 0$, we can have
$f_\infty \neq 0$, if $P(k=1)= 0$, $f_\infty = 0$.
Using Eq.(\ref{Eq.delta}), we
can calculate $\delta_A=\ln(3/2)/\ln(4/3)\approx 1.41$, 
$\delta_B=\ln(9/2)/\ln(20/9)\approx 1.88$ and $\delta_C=\infty$. It is clear that
network A and B have finite $\delta$, while for network C, $G^{'}_{1}(0)=0$ 
thus $\delta_c=\infty$.
In order to have finite $\delta$, 
$P(k=2)+P(k=1)$ must be greater than 0.
If $P(k=2)=P(k=1)=0$ (called the B$\ddot{o}$ttcher case \cite{bingham}),
then $\delta=\infty$,
which indicates that $\Phi(y)$ has an exponential singularity.
For the B$\ddot{o}$ttcher case, the distribution of $B_{\ell}$ is not described by a 
power law, i.e. there are no fractal boundaries.

\newpage
\vspace*{-0.3cm}

\begin{thebibliography}{99}
\bibitem{note1} Here $t$ represent the number of simulation steps of the
building process, each of which consists of selecting a single 
open link belonging
to the aggregate and connect it to another link (cases (i), (ii), (iii) 
shown in Fig.~\ref{fig1}).
\bibitem{footnote1}
If $r=r_l$ corresponding to a certain shell $\ell$, $r=\sum_k P_\ell(k)$
and $P_r(k)=P_\ell(k)/r$ in notations of Ref. \cite{kalisky}.
\bibitem{scatter} For different realizations from different root, 
the results shown in 
Fig.~\ref{fig3} and Fig.~\ref{fig4} are aligned nicely along the
same line. There are almost no deviations.
\end{thebibliography}
\end{document}